\documentclass[extra]{gji}
\usepackage{graphicx}
\usepackage{nicefrac}
\renewcommand{\vec}[1]{\mbox{\boldmath $#1$}}
\newcommand{\Om}{{\it{\Omega}}}

\renewcommand{\eps}{\textit{Earth~Planets~Space}}

\renewcommand{\gji}{\textit{Geophys.\ J.\ Int.}}

\renewcommand{\grl}{\textit{Geophys.\ Res.\ Lett.}}
\renewcommand{\jgr}{\textit{J.\ geophys.\ Res.}}
\renewcommand{\mnras}{\textit{Mon.\ Not.\ R.\ astr.\ Soc.}}
\renewcommand{\pepi}{\textit{Phys.\ Earth planet.\ Inter.}}

\newcommand{\gafd}{{\textit{Geophysical and Astrophysical Fluid Dynamics}}}
\newcommand{\apss}{\textit{Astrophysics and Space Science}}

\newcommand{\an}{\textit{Astronomische Nachrichten}}
\newcommand{\apj}{\textit{Astrophysical Journal}}
\newcommand{\aap}{\textit{Astronomy and Astrophysics}}
\newcommand{\pre}{\textit{Physical Review E}}
\begin{document}
\title[Anisotropic turbulence in rotating magnetoconvection]{Anisotropic
  turbulence in weakly stratified rotating magnetoconvection}
\author[A. Giesecke]{A. Giesecke$^1$\\
$^{1}$Astrophysikalisches Institut Potsdam, An der Sternwarte 16, D-14482 Potsdam, Germany
}
\maketitle
\begin{summary}
{
Numerical simulations of the 3D MHD-equations that describe rotating
magnetoconvection in a Cartesian box have been performed using the code NIRVANA.
The characteristics of averaged quantities like the turbulence intensity
and the turbulent heat flux that are caused by the combined action of the
small-scale fluctuations are computed.
The correlation length of the turbulence significantly depends on the strength
and orientation of the magnetic field and the anisotropic behavior of the
turbulence intensity induced by Coriolis and Lorentz force is considerably
more pronounced for faster rotation.
The development of isotropic behavior on the small scales -- as it is observed
in pure rotating convection -- vanishes even for a weak magnetic field which
results in a turbulent flow that is dominated by the vertical component.
In the presence of a
horizontal magnetic
field the vertical turbulent heat flux slightly increases with increasing field strength, so
that cooling of the rotating system is facilitated. 
Horizontal transport of heat is always directed westwards and towards the
poles. 
The latter might be a source of a large-scale meridional flow whereas the
first would be important in global simulations  in case of non-axisymmetric
boundary conditions for the heat flux.   
}
\end{summary}
\maketitle
\section{Introduction}

Anisotropic turbulence is a fundamental feature
in many astro- or geophysical systems.
Well known realizations are the turbulent motions in the solar convection zone 
or the flow of liquid iron in the outer core of the Earth.
Both kinds of convectively driven flows are assumed to be responsible for dynamo
action.
Flow and field are attuned in a complex non-linear system where the flow
produces the magnetic field which in turn backreacts on the field producing flow.
Direct observations of the sun offer the possibility to examine the details of the
complicated interactions between convection and magnetic fields. 
On the top of the solar convection zone, in the photosphere, where rotational
effects are rather unimportant, occasionally strong localized and radial oriented magnetic fields inhibit the convective transport
of energy which results in a cooler (and therefore darker) area, a sunspot
(see e.g. Weiss, 1990).
Numerical simulations of non-linear magnetoconvection
confirm that a vertical magnetic field suppresses
turbulent motions as well as it considerably reduces the size of the
convection cells, a result that has been known from the linear stability analysis
of Chandrasekhar (1961). 
The smaller cells come along with reduced variations of the
temperature so that the correlations of velocity and temperature fluctuations, the turbulent
heat flux, decrease with increasing field (Cattaneo et al., 2003).
Towards the edge of the sunspot the field changes its orientation to a more horizontal
direction.
As a result the
flow pattern changes to small brighter and darker
elongated filaments that surround the
sunspot (Weiss et al., 2004).
Thermodynamic properties like the heat transport and the temperature
distribution are undoubtedly linked to the anisotropy induced by the
dominant magnetic field component.
Closely related to the interaction of convection, magnetic field and turbulent
heat transport are questions about the existence of warmer (and therefore brighter)
rings around sunspots (Eschrich \& Krause, 1977; R\"udiger \& Kitchatinov, 2000) or the behavior of (star-) spots in fast rotating stars.
In contrast to the sun, observations of fast rotating stars using Doppler-Imaging
techniques reveal an accumulation of star spots close
to the poles (Strassmeier, 2002, 2006).
This phenomenon clearly indicates the different behavior of 
convectively driven turbulence in fast rotating objects and points out that a
different dynamo mechanism is running where differential rotation
is of minor importance (Bushby, 2003).
Comparable considerations might also be of interest for the Earth because
observations of the radial field component at the core
mantle boundary as well as highly resolved simulations of the geodynamo show
strong localized flux patches whose pattern and pair-like occurrence resemble
the behavior of sunspots (Roberts \& Glatzmaier, 2000; Jackson, 2003).

However, the particular properties of a fast rotating planetary body like the Earth lead to physical
conditions that differ significantly from the sun or stellar objects.
A small density contrast between the top and the bottom of the fluid outer
core and a nearly incompressible fluid result in a small Mach number flow
which is essentially influenced by the non-linear back-reaction of a strong
magnetic field.
Convection in the fluid outer core is mainly affected by two dominant forces, the Lorentz force due to the strong
magnetic field and the Coriolis force due to the fast
rotation of the Earth.
In such a rapidly rotating convection driven dynamo the turbulence is subject to
three preferred directions defined by magnetic field, rotation axis
and gravity. 
The arising
anisotropies result in a plate like form of convection
cells which are aligned along the
rotation axis and elongated in the direction of the dominant magnetic field
component (Braginsky \& Meytlis, 1990; St. Pierre, 1996; Matsushima et al., 1999).  

Despite the progress in computational and numerical techniques that have been made
during the recent years, it is still impossible to simulate turbulent
convection in rotating spherical shells
with a sufficient resolution so that the resolved scale range of the turbulence is
rather restricted (Hollerbach, 2003).
Furthermore, anisotropy usually is neglected by an implicit
assumption of a large eddy simulation where 
scalar parameters resemble the turbulent values of
the diffusivities.
Beside the fact that mostly the diffusivities -- for the reason of numerical
stability -- have to be chosen much larger than
even the (poorly known) ``real'' turbulent values, this oversimplifying
assumption describes an isotropic transport of flux and an isotropic
dissipation of energy.
This is only justified if either the resolution is high enough so that all
dynamical important modes are resolved or if no preferred direction originated by
external forces, stratification or boundary conditions exists. Both, in general,
are not the case. 
In order to include anisotropic effects in a more sophisticated way, turbulence
models have to be considered where tensorial expressions for the
diffusivities take
care of the directional dependence of the turbulence and the influence of
the unresolved scales is parameterized in terms of the resolved large-scale fields.

Tensor coefficients for the viscous and thermal diffusivity of a fast rotating
and a strong field model of the Earth's core have been derived by
Phillips \& Ivers (2001, 2003).
They present expressions that describe enhanced diffusion
along a dominant  azimuthal magnetic field. 
Non-diagonal elements are neglected so that diffusion in the horizontal
directions due to a coupling between vertical and horizontal
components of the turbulent flow is not included.
The results are intended for the use in
pseudo-spectral codes but at this stage no
applications in numerical investigations are available so that the
consequences for global simulations remain unknown.

A different approach is examined by Buffett (2003) and Matsushima (2004, 2005). 
They present a subgridscale (SGS) model which is essentially based  on the self-similarity of the
turbulence.
By comparison of a highly resolved direct numerical simulation
(DNS) with a large eddy simulation based on the SGS model, they show
that the SGS model is able to predict appropriate anisotropic heat- and momentum fluxes.
An advanced version of the SGS model that
includes the Lorentz- and induction terms is introduced by Matsui \& Buffett (2005).
They apply the so called non-linear gradient model, an
adaption of the similarity model that additionally is based on the local character of the turbulence.
In general, the SGS models coincide rather well with the direct numerical
simulations, however, both models are not able to reproduce all details of the spatiotemporal behavior.
Furthermore, systematic deviations are obtained in mean quantities, like kinetic and magnetic energy,
which are situated between the values from the resolved DNS and the values of a truncated, unresolved DNS.

In order to evaluate the functionality of SGS models it is
useful to  further investigate the development of a convectively
driven turbulence in view of
the dependence on field strength and latitude.
Subject matter of the present paper is the connection between turbulence
characteristics, that are caused by the
average action of the small-scale fluctuations 
and the turbulent transport of heat which may essentially contribute to the
conditions that determine the
temperature distribution within the fluid outer core.
A simplified Cartesian system is examined where a convectively driven turbulence
in a conducting fluid is subject to fast rotation and an externally imposed magnetic field.
As we do not intend to reproduce
geophysical realities in all details, the presented investigation is based on a highly idealized system where the
emphasis lies on the anisotropy inducing effects of rotation and magnetic field. 
Properties like the exact behavior of the temperature
gradient, the equation of state or the influence of curvature are assumed to
be of secondary importance.
\section{The model}
\subsection{General properties}
\begin{figure}
\includegraphics[width=8.5cm]{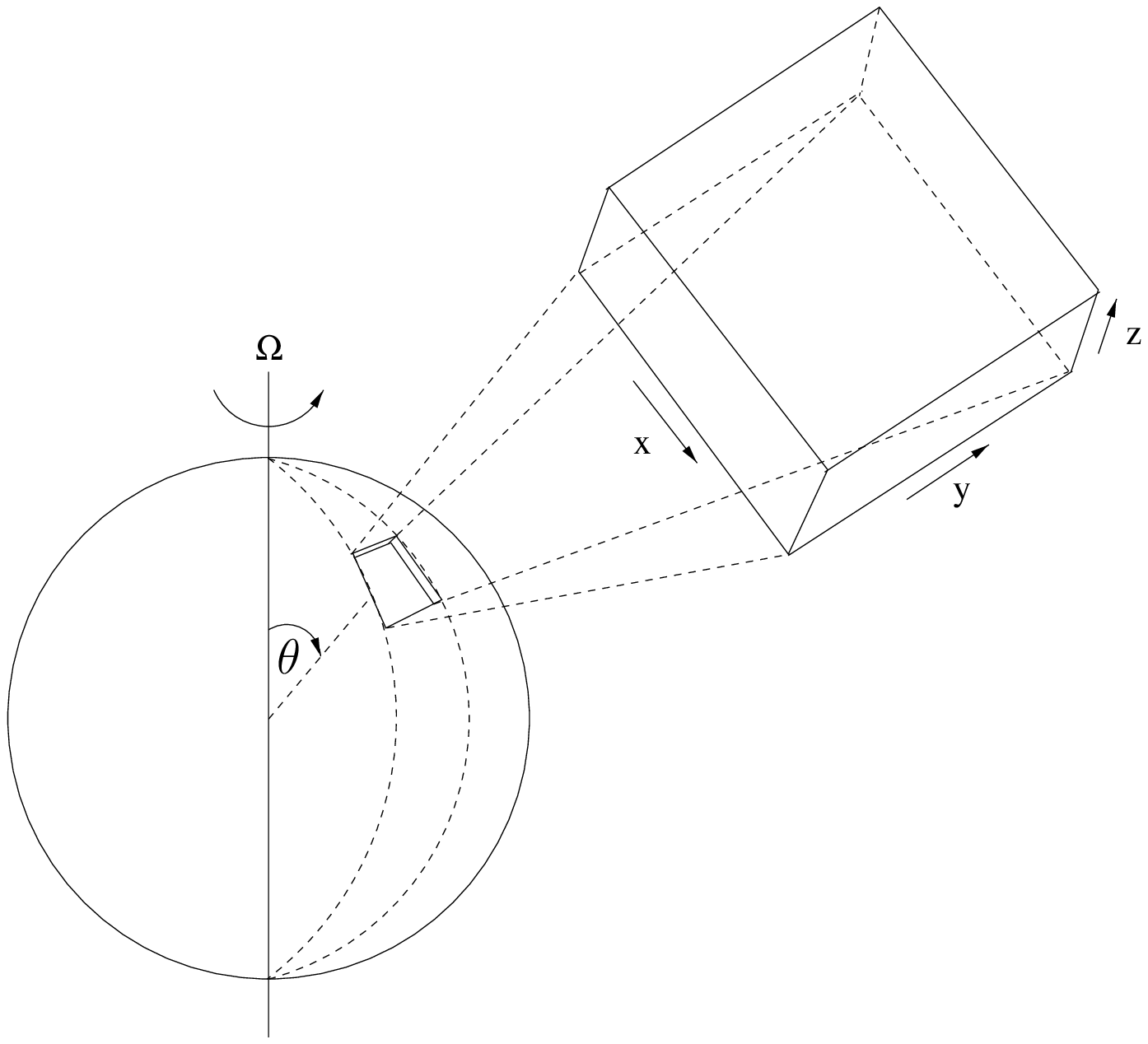}
\caption{ Model box as part of a rotating spherical shell at co-latitude
  $\theta$.} \label{modell_skizze}
\end{figure}

A detailed description of the applied local model can be found in
Giesecke et al. (2005). 
Figure~\ref{modell_skizze} shows a sketch of the computational domain, a Cartesian box placed somewhere on a
spherical shell at a co-latitude $\theta$.
The unit vectors ${\vec{
\hat{x},\hat{y},\hat{z}}}$ form a
right-handed co-rotating  Cartesian coordinate system with $\vec{\hat{x}}$ pointing towards the
equator, $\vec{\hat{y}}$ pointing in the toroidal direction (from west to east) and $\vec{\hat{z}}$ pointing from the
bottom to the top of the box.
In global spherical coordinates, $\vec{\hat{z}}$
represents the radial direction $\vec{\hat{r}}$ (oriented from inside to
outside), $\vec{\hat{y}}$ the azimuthal direction $\vec{\hat{\varphi}}$
(oriented eastwards) and $\vec{\hat{x}}$ the
meridional direction $\vec{\hat{\theta}}$ (oriented towards the equator).
The angular velocity  $\vec{{\it{\Omega}}}$ in the co-rotating local box coordinate system 
is given by
$\vec{{\it{\Omega}}}=-{\it{\Omega}}\sin\theta\hat{\vec{x}}+{\it{\Omega}}\cos\theta\hat{\vec{z}}$
where ${\it{\Omega}}$ is the angular velocity of the rotating spherical
shell.
The box with an aspect ratio 8:8:1 is placed at different 
latitudes  on the northern hemisphere of the rotating spherical shell and a standard resolution of $100\times100\times80$ grid points is used in all calculations. 
The
co-latitude angle $\theta$ is varied from $\theta=0^{\circ}$ (north pole) to $\theta=75^{\circ}$.
\subsection{Equations}
The MHD-equations for a rotating fluid, including
the effects of thermal conduction, compressibility, viscous friction and
losses due to magnetic diffusivity, are
\begin{eqnarray}
\partial_t\rho&=&-\nabla\cdot(\rho\vec u), \label{conteq}\\
\partial_t(\rho\vec{u})&=&-\nabla\cdot(\rho\vec{uu})-\nabla p+\nabla\cdot\sigma+\rho\vec{g}\label{nseq}\\
&&+\frac{1}{\mu_0}(\nabla\times\vec{B}\mathrm)\times\vec{B}-2\rho\vec{\Om}\times\vec{u},\nonumber\\
\partial_t e&=&-\nabla\!\cdot\!(e\vec{u})\!-\!p\nabla\!\cdot\!\vec{u}\!+\!\sigma
\!\cdot\!\nabla\vec{u}\!+\!\frac{\eta}{\mu_0}|\nabla\times\vec{B}|^2\label{eneq}\\
&&+\nabla\!\cdot\!(\kappa\nabla T),\nonumber\\
\partial_t\vec{B}&=&\nabla\times(\vec{u} \times
\vec{B}-{\eta}\nabla\times\vec{B}).\label{indeq}
\end{eqnarray}
Here, $\rho$ denotes the density, $\vec{u}$ the velocity, $p$ the pressure,
 $\vec{B}$ the magnetic flux density, $T$ the temperature and $e$ the thermal energy density. 
A constant gravitational field
$\vec{g}=-g\hat{\vec{z}}$ is assumed within  the domain.
The viscous stress tensor $\sigma$ is given by
$\sigma_{ij}=\nu\rho\left(\partial_j{u}_{i}+\partial_i{u}_{j}-\nicefrac{2}{3}\nabla\cdot\vec{u}\delta_{ij}\right)$.
$\nu$ denotes the kinematic viscosity and $\kappa$ the thermal
conductivity coefficient.
The values of $\kappa$, the dynamic viscosity 
$\nu_{\mathrm{dyn}}=\nu\rho$ and the magnetic diffusivity $\eta$ are constant within the box volume.
An ideal gas equation of state is assumed:
\begin{equation}
p=(\gamma-1)e=\frac{k}{m\bar \mu}\rho T
\label{idealgaseq}
\end{equation}
where $k$ is the Boltzmann constant, $m$ the atomic mass unit, 
$\bar\mu$ the
mean molecular weight ($\bar{\mu}=1$ for all runs) and
$\gamma=c_p/c_V=5/3$ is the ratio
of  $c_{p}$, the specific heat at constant pressure, and $c_V$, the specific heat at
constant volume.
The permeability $\mu_0$ is given by the vacuum value $\mu_0=4\pi\times 10^{-7}\mathrm{VsA^{-1}m^{-1}}$.

The equations (1) -- (4) are solved applying the code
NIRVANA (Ziegler 1998, 1999). 
NIRVANA makes use of a dimensional- and operator splitting approach. 
A second-order accurate finite-volume scheme with a piecewise linear
reconstruction and monotonized slope limiter (van Leer) is used for the hydrodynamic
advection part of the equations.  A constraint transport solver utilizing the
method of characteristics is employed for the numerical solution of the induction equation. 
The time integration of the source terms is performed by an explicit
Euler scheme (except
the Coriolis force which is treated analytically).
\subsection{Initial state and input parameters}
The simulations are started with an initial state that is determined by a
hydrostatic equilibrium  $\partial_z p = -\rho g$ and a polytropic temperature
distribution $T=T_0\left({\rho}/{\rho_0}\right)^{\Gamma}$ in the
absence of motions (the subscript 0 refers to values taken at the top boundary of the box).
With the equation of state (\ref{idealgaseq}) the initial density distribution is given by
\begin{equation}
\rho(z) = \rho_0\left(1+\frac{dT/dz}{T_0}(d-z)\right)^{1/\Gamma},
\label{eq_rhodistribution}
\end{equation}
where $d$ stands for the vertical box extension and the polytropic index $\Gamma$
is given by
$\Gamma=\ln \left(1+d\frac{dT}{dz}/T_0\right)/\ln \xi$.
The stratification index $\xi =\rho_{\mathrm{bot}}/\rho_{\mathrm{0}}$,
the temperature $T_{\mathrm{0}}$ and the global temperature gradient $dT/dz$ are prescribed
input parameters.

The parameters $\nu, \kappa$ and $\eta$ are calculated from the Rayleigh
number Ra,
\begin{equation}
{\rm Ra}=\frac{{c_{{p}}}d^4}{\kappa\nu}\frac{\rho g}{T}\left(\frac{dT}{dz}-\frac{g}{c_p}\right)
\label{gg}
\end{equation}
 with $c_{p}={k}({m\bar{\mu}})^{-1}\gamma(\gamma-1)^{-1}$ the specific heat at
constant pressure, the Prandtl number ${\rm Pr}=\nu\rho c_{{p}}/\kappa$ and
the magnetic Prandtl number ${\rm Pm}=\nu/\eta$.  
The basic parameter set used for all
simulations that are presented in this paper is given by ${\rm Ra}=10^6, {\rm
Pr}=0.5$ and ${\rm Pm}=0.5$.
The diffusivity parameters are scalar quantities assuming that in the local
model the resolved scales contain most of the energy, so that the anisotropic transport is
described with sufficient accuracy and the remaining non-resolved modes are
less important.
The rotation rate ${\Om}$ is
parameterized by the Taylor number given by 
\begin{equation}
{\rm Ta}=\frac{4{\Om}^2d^4}{\nu^2}
\end{equation}
which is related to the
Ekman number by $\mathrm{Ta}=\mathrm{Ek}^{-2}$. 
Essentially, a rotating system is examined where $\mathrm{Ta}$ is set to a
value of $10^7$.
A few simulations with a Taylor number $\mathrm{Ta}=10^6$ have been performed
which demonstrate the significant changes in the behavior of the turbulence
as the rotation rate increases.
The magnetic field is
expressed by the Els\"asser number:
\begin{equation}
\Lambda={\vec{B}^2\over{2\Om\mu_0\rho\eta}}.
\label{els}
\end{equation}
$\Lambda$ represents the relation of the Lorentz force to the Coriolis force,
which are assumed to be of the same order of
magnitude within the fluid outer core. 
The original picture favors $\Lambda\approx O(1)$ because at this
value the critical Rayleigh number $Ra_{\mathrm{crit}}$ at which the onset of thermal
convection occurs, becomes minimal.  
In that case the Coriolis force can be balanced by the Lorentz force.
Then, there remains no need for 
balancing these terms with the (extremely small) viscous terms which would force the
convection to occur on very short length scales. 
Such a flow would be much more difficult to maintain which is evident on the
basis of the suppression of convective motions in simple rotating convection or non-rotating
magnetoconvection. 
For a detailed description of the underlying mechanism see e.g. R\"udiger \&
Hollerbach (2004).
However, Zhang \& Jones (1994) did not find such a minimum for convection in a
rotating spherical shell, but  they were able to show that the existence of a
stable magnetic field requires an Els\"asser number in the range $1\la \Lambda
\la 10$.

For the comparison of the magnetic quenching character in simulations with different rotation rates, $\Lambda$ is not a suitable parameterization of the field strength.
For that case, the dependence on the imposed field strength is expressed in units
of the equipartition field strength $B_{\mathrm{eq}}$ defined by
\begin{equation}
B_{\mathrm{eq}}=\sqrt{\mu_0\rho}u_{\mathrm{rms}},
\end{equation}
where $u_{\mathrm{rms}}$ denotes the root-mean-square velocity (see definition in  the following section)
for non-rotating and non-magnetic convection (in code units: $u_{\rm{rms}}\approx 19.5$).
%
%
\subsection{Averaging procedure}
Due to the small stratification all
considered quantities do only weakly depend on the vertical coordinate $z$ (except close to the upper and the lower
boundaries). Therefore it is justified to characterize the
turbulence properties of a fluctuating quantity by a volume average given by
\begin{equation}
{\left<{f'}\right>}=\frac{1}{N} \sum_{i,j,k}
\underbrace{\Bigl(f_{i,j,k}(t)-\overline{f}_z(t)\Bigr)}_{\displaystyle f'_{i,j,k}}.
\end{equation}
On the left hand side, $f'$ represents a fluctuating quantity and
the brackets $\langle\cdot\rangle$ denote the averaging procedure.  
On the right hand side, $f_{i,j,k}(t)$ represents the numerically computed value of the
quantity $f$ at a certain grid cell labeled by $i,j,k$ at a certain time $t$.
$\overline{f}_z(t)$ represents the horizontal average of the considered quantity.
$N=n_xn_yn_z$ denotes the
number of grid cells used for averaging.
Time averaging is done for periods with no significant changes in the
statistically steady convection state.
Convergence of the time-averaged solutions
has been checked by comparing results obtained from averages over different
(increasing) periods. 
Reasonable results are obtained for averaging periods larger than at
least 20 turnover times $\tau_{\mathrm{adv}}=d/u_{\mathrm{rms}}$, where the
root mean square velocity $u_{\mathrm{rms}}$ is defined as
$u_{\mathrm{rms}}={\sqrt{\langle {\vec{u'}^2}\rangle}}$.

\medskip

All quantities show fluctuations in time with a standard deviation of the order of 10 percent of their
mean value. These fluctuations become smaller for increasing field strength,
but within the context of this paper they do not exhibit any essential
features so that the standard deviations are omitted for the reason of clarity.
\subsection{Boundary Conditions}
All quantities are subject to periodic boundary conditions in the horizontal directions.
At the top and at the bottom of the computational
domain constant values for density and temperature are imposed.
The vertical boundary condition for the magnetic field is a perfect conductor condition, and a stress-free
boundary condition is adopted for the horizontal components of the velocity $u_x$ and $u_y$.
Impermeable box walls at the top and the bottom lead to a vanishing
$u_z$ at the vertical boundaries. 
Table~\ref{bc_tab} summarizes the boundary conditions and gives the
initial values for density and temperature which describe the overall stratification and the global temperature
gradient.
\begin{table}
\caption{Vertical boundary conditions}
\begin{tabular}{lllrr}
\hline
 & $\rho$  & $T$ & $\vec{u}$ &
$\vec{B}$
\\
\hline
{ top} & $1$ & $1$ & $\partial_z u_x = 0$ & $\partial_z B_x = 0$
\\
 $(z=d)$ & & & $\partial_z u_y = 0$ &  $\partial_z B_y = 0$
\\ 
& &  & $u_{z} = 0$ & $B_{z} = 0$
\\
\hline
\hline
{bottom}  & $1.1$ & $2$ & $\partial_z u_x = 0$ & $\partial_z B_x = 0$
\\
$(z=0)$ & & & $\partial_z u_y = 0$ & $\partial_z B_y = 0$
\\ 
& & & $u_{z} = 0$ & $B_{z} = 0$
\\
\hline
\end{tabular}
\label{bc_tab}
\end{table}
For all simulations, temperature and density at the top of the box are
scaled to unity, as it is the case for the
global temperature gradient $dT/dz$ and the box height $d$. 
A stratification index of $\xi=\rho_{\mathrm{bot}}/\rho_{\mathrm{0}}=1.1$ is used.

The boundary conditions certainly influence the behavior of the flow within
the computational domain. To reduce the undesired effects of the boundaries all volume
averages are performed over the inner part of the computational domain (between
$z=0.2$ to $0.8$). 
Naturally, such a simplifying box model is not able to represent
the large scale flows that may occur in a rotating sphere. 
Especially the existence of periodic boundary conditions in the horizontal
directions filters out the fraction of the solution with wavelengths larger
than the horizontal extension of the box (of course, this behavior is desired,
because here only the small scale fluctuations are of interest).
However, such filtering is not possible in the vertical
direction due to the existence of density stratification and temperature gradient.
\section{Results}
\subsection{Convection pattern and anisotropy}
\begin{figure*}
\hspace*{-0.8cm}
\includegraphics[width=10.5cm,angle=0]{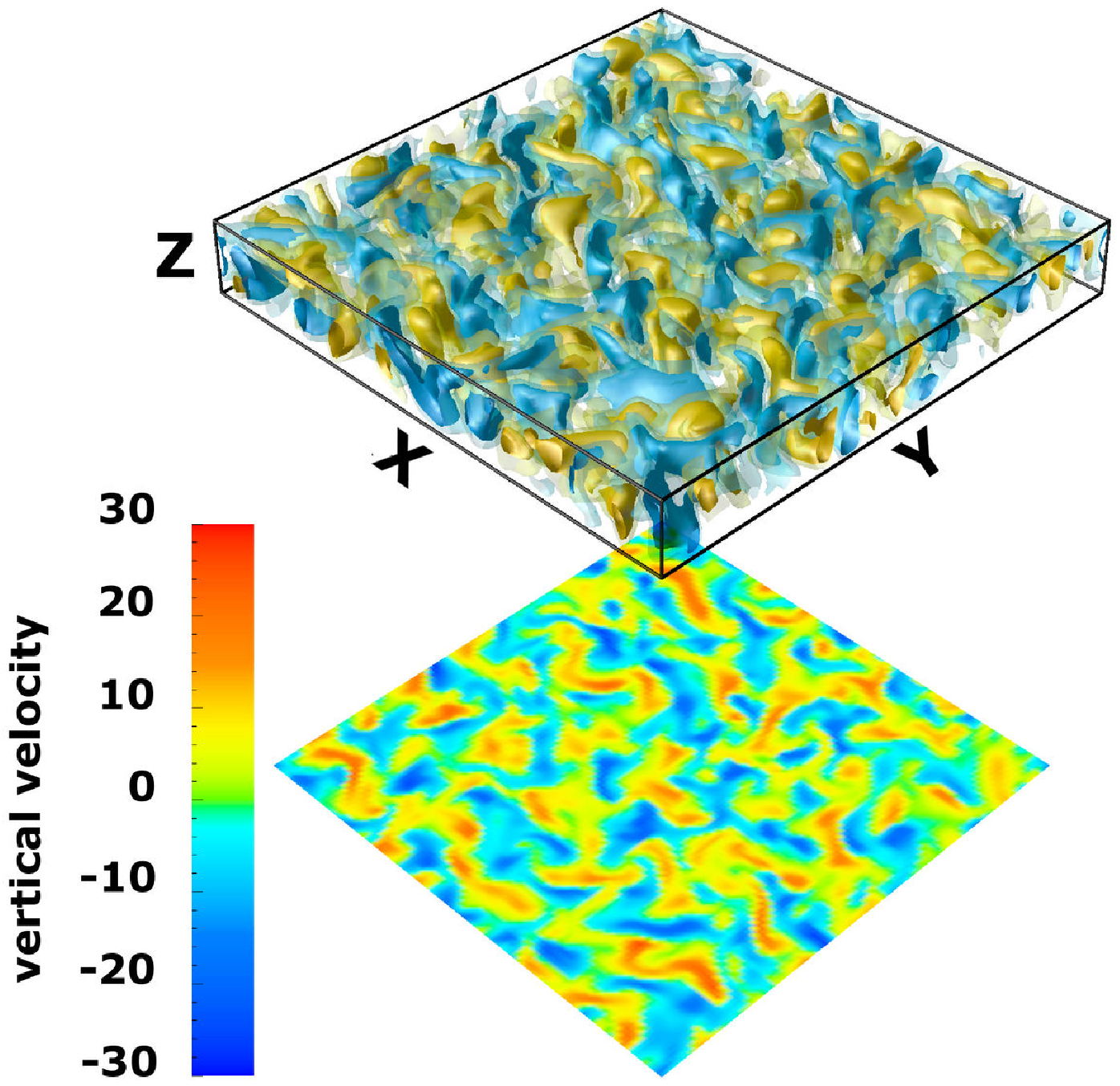}\nolinebreak[4!]
\hspace*{-1cm}
\includegraphics[width=10.5cm,angle=0]{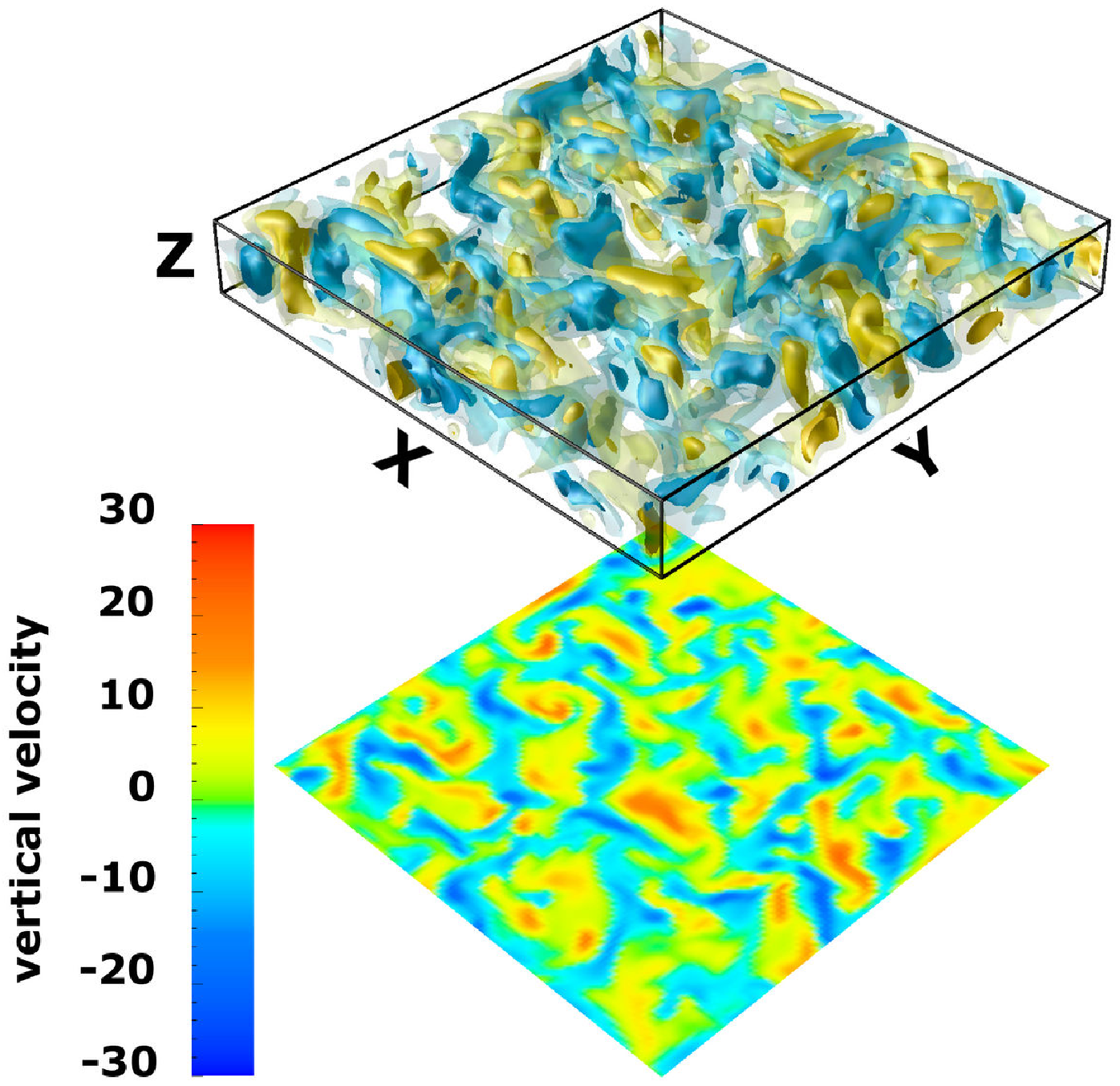}\\
\hspace*{-0.8cm}
\includegraphics[width=10.5cm,angle=0]{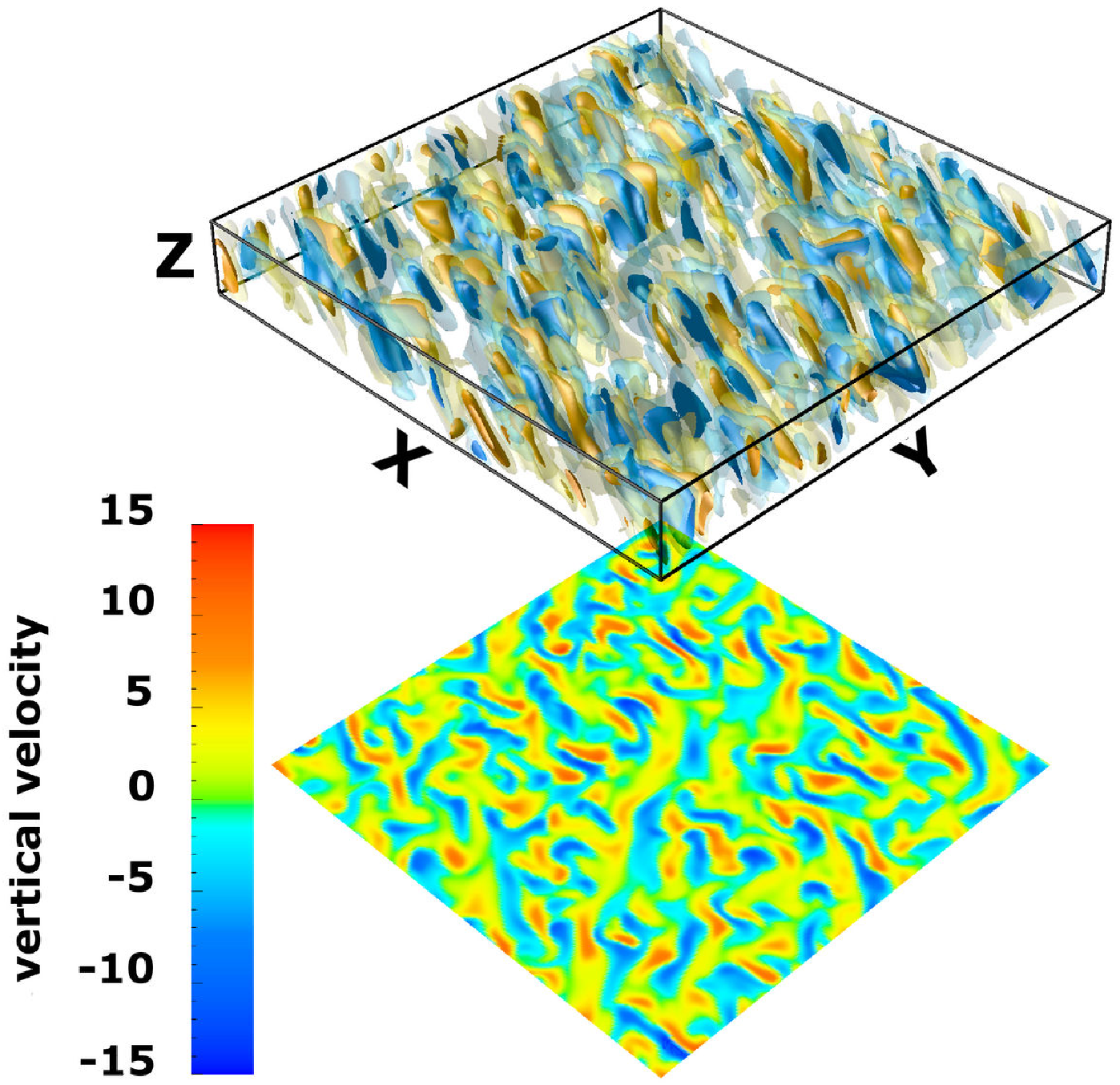}\nolinebreak[4!]
\hspace*{-1cm}
\includegraphics[width=10.5cm,angle=0]{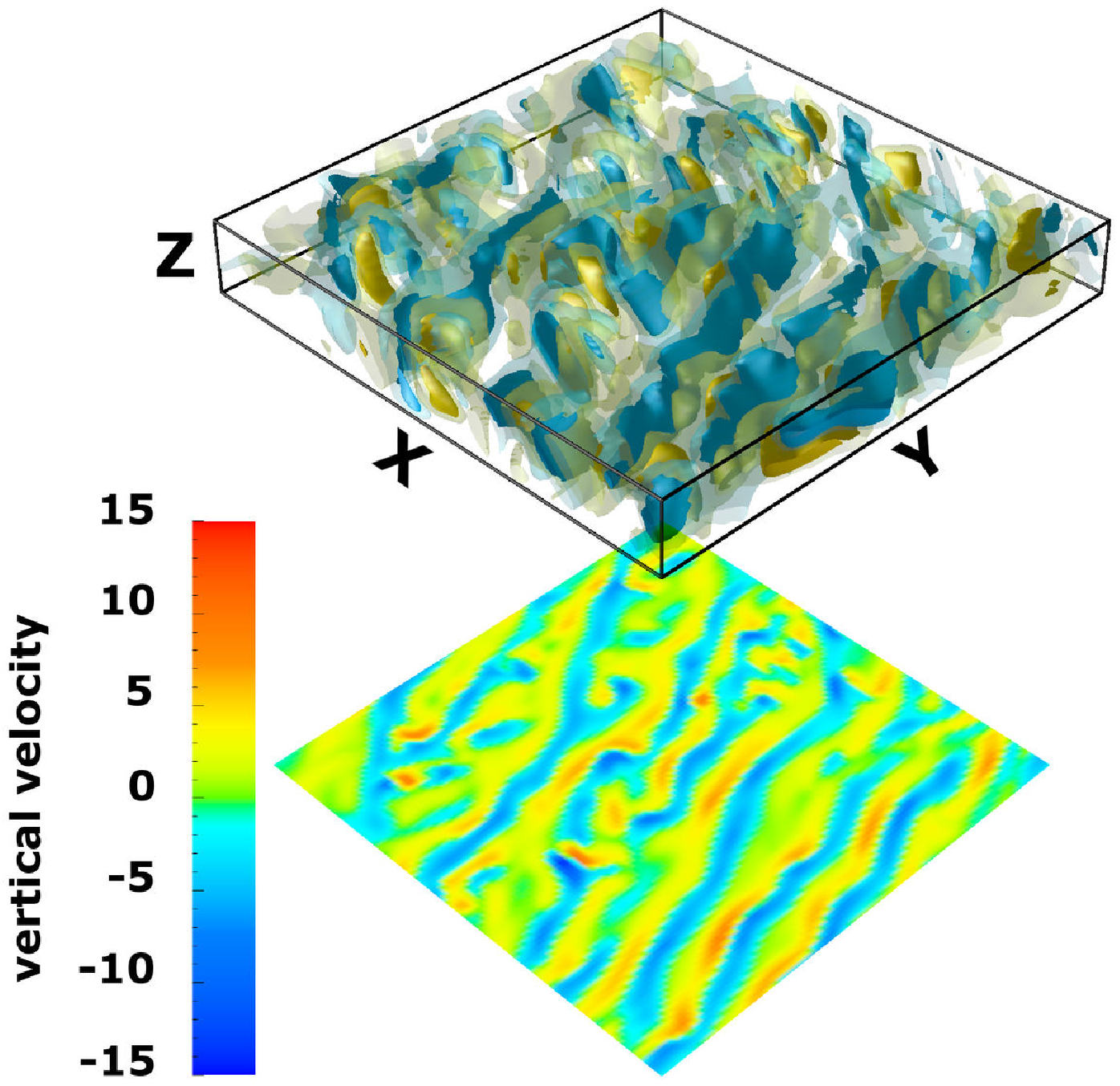}
\caption{$z$-component of the velocity for ${\rm Ta}=10^6$ (upper row) and ${\rm Ta}=10^7$ (lower row). Left side:
  $\Lambda=1$, right side: $\Lambda=4$. All plots present a snapshot at $\theta
  = 45^{\circ}$. The solid iso-surfaces denote the vertical velocity at
  $u_z=u_{\rm{rms}}$, and the
transparent iso-surfaces denote the pattern at
  $u_z=0.5u_{\rm{rms}}$ (upper left $u_{\rm{rms}}\approx 11.5$, upper right:
  $u_{\rm{rms}}\approx 10.8$, lower left: $u_{\rm{rms}}\approx 5.4$, lower right:
  $u_{\rm{rms}}\approx 4.7$, all numbers in code units). A horizontal cut of the velocity field at a depth $z=0.5$ is projected on a
plane below the three dimensional box. Blue (dark) tones represent downwards oriented flow and yellow/red (light) colors represent
upwards oriented flow.
}\label{vz_pattern}
\end{figure*}

Within the applied parameter regime all simulations  result in a pressure dominated ($\beta=2\mu_0p/B^2 \gg 1$) low Mach number
flow ($\mathrm{Ma}=u_{\mathrm{rms}}/c_{\mathrm{s}}\ll 1$, with the sound
speed $c_{\mathrm{s}}=\sqrt{\nicefrac{kT_0}{m\overline{\mu}}}$).
Depending on rotation rate and field strength $\mathrm{Ma}$
is of the order of $10^{-2}$ (e.g. for $\Lambda=1$ and
$\mathrm{Ta}=10^7$ the characteristic Mach number is given by $\mathrm{Ma}\approx0.03$).
The corresponding Reynolds numbers $\mathrm{Re}=u_{\mathrm{rms}}d/\nu$ are of
the order of 100.
The results that are presented in the following are obtained from simulations with an imposed horizontal field
($B_y$) which corresponds to a toroidal field in spherical coordinates. 
The typical structures of the convective motions are visualized in Fig.~\ref{vz_pattern}. 
The upper row presents the vertical velocity pattern obtained from a
simulation run with $\mathrm{Ta}=10^6$ (left: $\Lambda\approx 1$ respectively
$B=0.1B_{\mathrm{eq}}$, right: $\Lambda\approx 4$, respectively
$B=0.2B_{\mathrm{eq}}$).
The lower row presents the case $\mathrm{Ta}=10^7$ (left: $\Lambda\approx 1$ respectively
$B=0.2B_{\mathrm{eq}}$, right: $\Lambda\approx 4$, respectively
$B=0.4B_{\mathrm{eq}}$).
All plots show a time snapshot of the vertical velocity pattern from a simulation
run at $\theta=45^{\circ}$.

The cell-like objects represented by the volume
rendered iso-surfaces indicate coherent large-scale
structures which span the full vertical extend of the domain. 
These convection cells are
continously formed, rearranged  and dissolved leading to a quasi-stationary
turbulent pattern which is more irregular in the interior of the domain.
According to the Taylor-Proudman theorem the convection cells are aligned with the rotation axis.
The average horizontal size of the cells strongly depends on
the rotation rate as well as the magnetic field strength and direction. 
The cell size is reduced for an increasing rotation rate and
if the horizontal magnetic field exceeds a certain threshold the
cells become elongated in direction of the dominant magnetic field component.
With respect to Fig.~\ref{vz_pattern} this is only noticeable in the faster
rotating case whereas for ${\mathrm{Ta}}=10^6$
the convection pattern remains nearly unaffected as the Els\"asser number  is
increased from $\Lambda\approx 1$ to $\Lambda\approx 4$. 

Opposite to a configuration with a large density contrast 
where the structure of the convection pattern 
consists of isolated, broad warm upflows and narrow network-like, cold and strong downflows,
within a weakly stratified layer the separation of up- and downflows is less pronounced.
Instead the upflows form a kind of
unconnected single plume-like structures enclosed by a loosely connected broad
network of downflows.
Up- and downflows are of approximately equal amplitude and occupy roughly an equal area in the
horizontal plane.

To characterize the anisotropy of the turbulence and to quantify the visual
impression obtained from Fig.~\ref{vz_pattern} the average extension of a convection
cell is estimated in both horizontal directions. 
The two-point correlation function $Q_{zz}$ is defined for the direction perpendicular to the
imposed magnetic field by
\begin{equation}
Q_{zz}(\delta x)=\frac{\left<u'_z(x)u'_z(x+\delta
    x)\right>}{\left<u'_z(x)^2\right>}
\end{equation}
and for the direction parallel to the field:
\begin{equation}
Q_{zz}(\delta
    y)=\frac{\left<u'_z(y)u'_z(y+\delta y)\right>}{\left<u'_z(y)^2\right>}.
\label{corfunc}
\end{equation}
$Q_{zz}$ can be used as a convenient measure for the average
horizontal extension of a convection cell.
Motions within a cell
are oriented in the same direction so that they are highly correlated whereas
no correlations exist if the distance between the considered two points (given by
$\delta x$, respectively $\delta y$) exceeds the size of the cell. 
The mean cell size is interpreted as the correlation length $\lambda_{\mathrm{corr}}$
or Taylor microscale, which is the characteristic length-scale of the
vorticity filaments observed in swirling flows. 
$\lambda_{\mathrm{corr}}$ is estimated from a function $f$, defined as 
\begin{equation}
f(\delta x) = 1-\lambda_{{\rm{corr}}}^{-2}(\delta x)^2
\label{taylor_eq}
\end{equation}
which is adjusted to the decreasing part of
the two-point correlation function $Q_{zz}$.

\begin{figure}
\includegraphics[width=8.5cm]{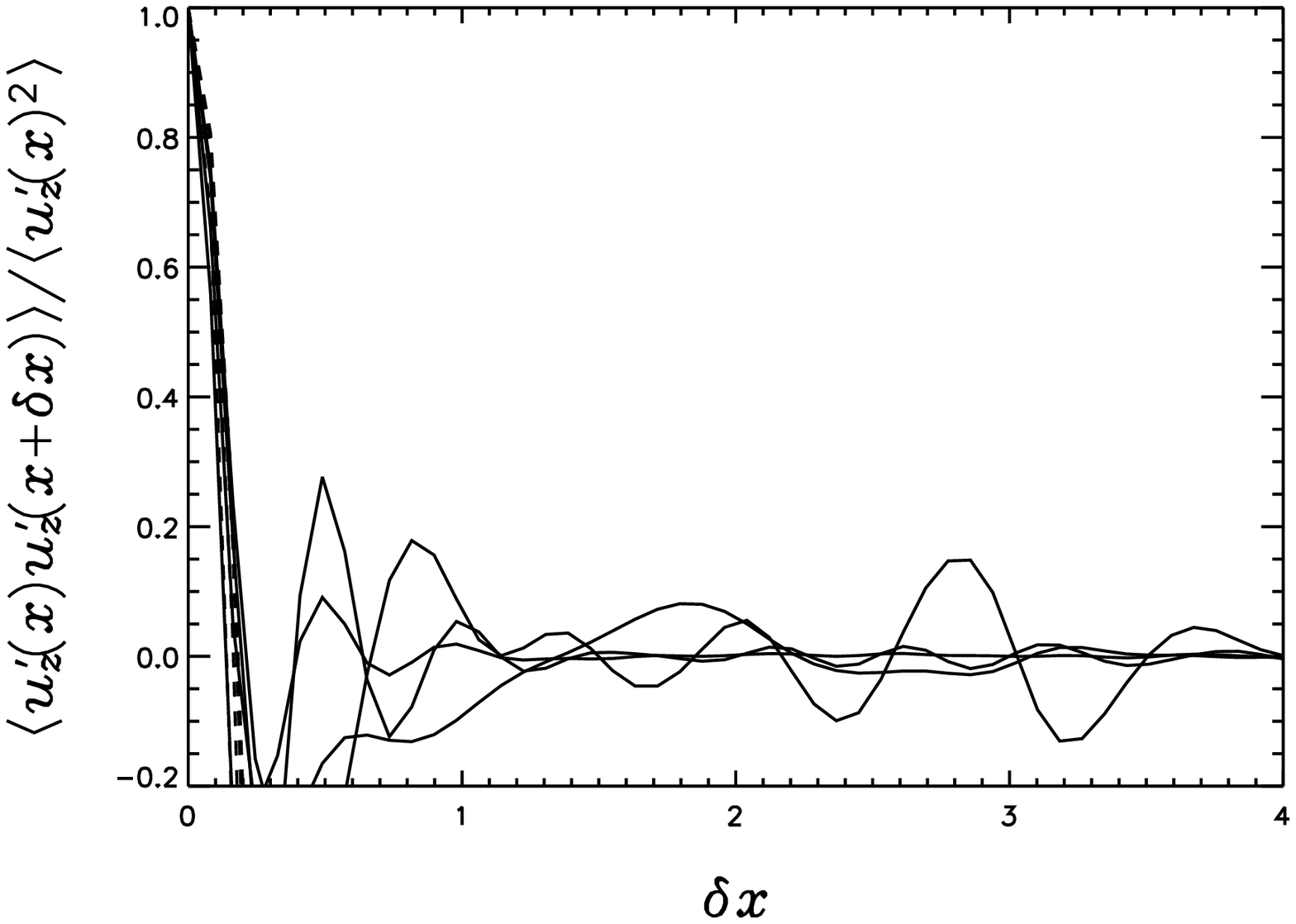}
\\
\includegraphics[width=8.5cm]{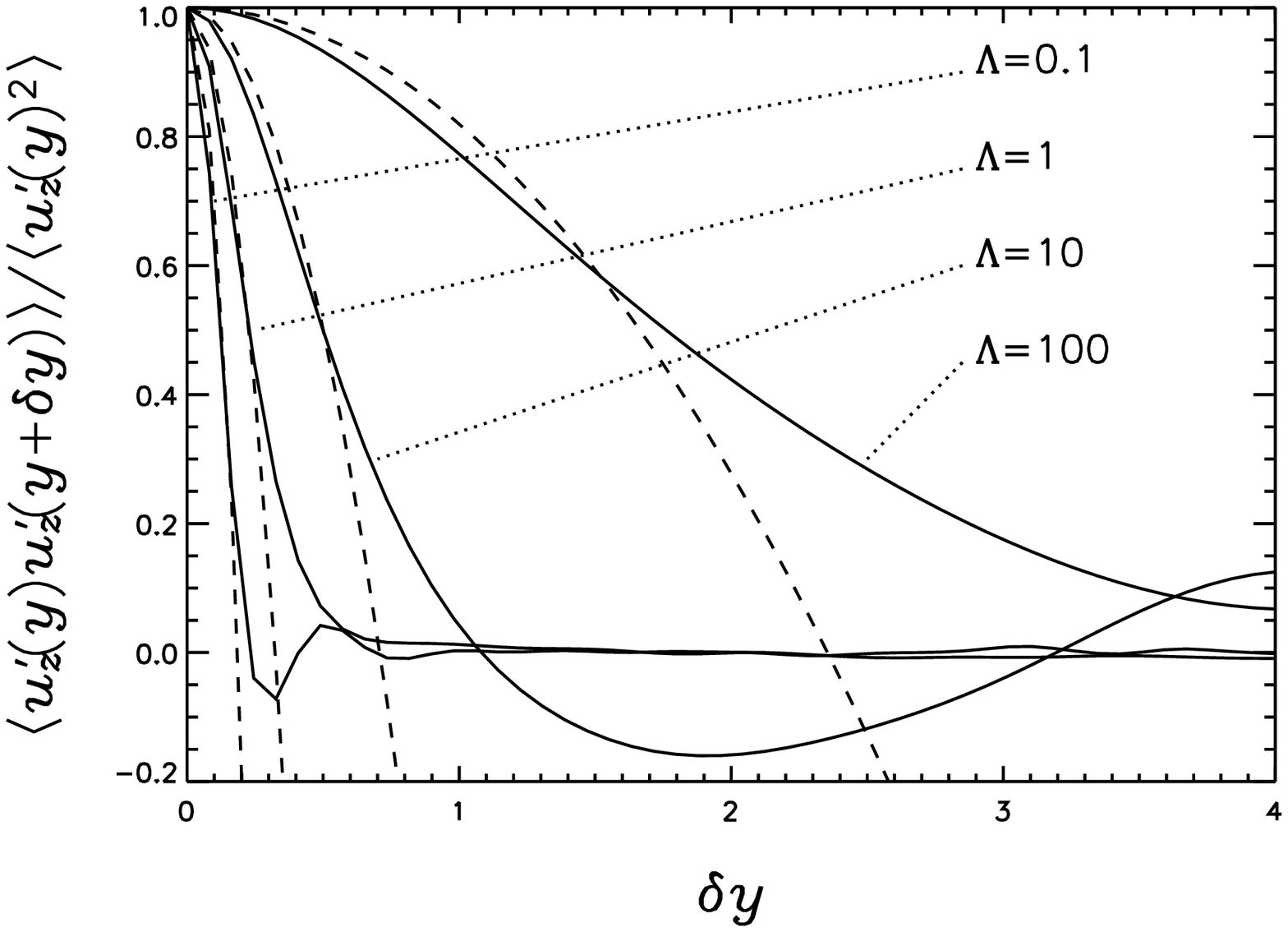}
\caption{Time average of the two-point-correlation functions. The solid lines represent
  $Q_{zz}(\delta x)$, respectively $Q_{zz}(\delta y)$ and the dashed lines represent a
  fit according to Eq.~(\ref{taylor_eq}) to the innermost part of $Q_{zz}$. $\mathrm{Ta}=10^7$, $\theta=0^{\circ}$, $\Lambda\approx
  0.1, 1, 10, 100$.}
\label{corfunc_plot}
\end{figure}

Figure~\ref{corfunc_plot} shows $Q_{zz}(\delta x)$ (upper panel)
and $Q_{zz}(\delta y)$ (lower panel) for $\Lambda=0.1,1,10,100$ (solid curves)
and the corresponding functions $f$ (dashed curves).
The shape of the curves of $Q_{zz}$ resembles the decrease of the correlation of
the turbulent vertical velocity at two different coordinates with
increasing distance between these two points.
The behavior of $Q_{zz}(\delta x)$ which determines the
correlation length perpendicular to $B_{y}$ is independent of the imposed field strength (therefore the
curves in the upper panel of Fig.~\ref{corfunc_plot} are not labeled by
$\Lambda$).
However, for large $\delta x$ the curves in the upper panel exhibit an
oscillating-like structure, which evolves in dependence of the field strength. 
With increasing field strength the amplitude increases and the length scale of
these oscillations decreases.
In case of weak fields the correlations outside of one cell are averaged out
due to the disordered occurrence of convection cells.
For strong fields the sheetlike structures that can be observed on the lower
right panel of Fig. 2 establish a more ordered structure which resembles in
the oscillations of the correlation function in the upper panel of Fig.~\ref{corfunc_plot}.

\begin{figure}
\includegraphics[width=8.5cm]{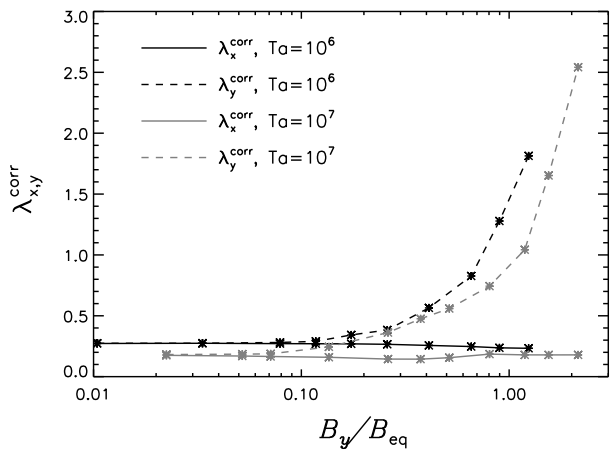}
\caption{Time and volume average of the correlation length $\lambda^{\mathrm{corr}}_{x,y}$ in $x$- and
  $y$-direction in dependence of the imposed magnetic field strength. The solid (dotted) line denotes the extension in
  $x$-direction ($y$-direction) for $\mathrm{Ta}=10^6$ (black) and for
  $\mathrm{Ta}=10^7$ (gray). $\theta=0^{\circ}$.  
}
\label{corlength}
\end{figure}

The influence of the magnetic field on the cell structure is considerably pronounced
in the lower panel of Fig.~\ref{corfunc_plot} where $Q_{zz}$ is broadened with increasing field strength.
The resulting correlation lengths $\lambda^{\mathrm{corr}}_{x,y}$ -- estimated 
independently for both horizontal directions from Eq.~(\ref{taylor_eq}) -- are shown in Fig.~\ref{corlength}.
The black curve denotes the case $\mathrm{Ta}=10^6$ where the solid
(dotted) line represents $\lambda^{\mathrm{corr}}_x$
($\lambda^{\mathrm{corr}}_y$). The gray curve shows the same quantities for $\mathrm{Ta}=10^7$.
As already indicated in the upper panel of Fig.~\ref{corfunc_plot}, for both
rotation rates, $\lambda^{\mathrm{corr}}_x$ is
nearly independent from the field strength so that the extension of the cell in
$x$-direction is only determined by the rotation
rate.
According to the linear theory the preferred length scale for the onset of
rotating convection scales as $\mathrm{Ta}^{-1/6}$ (for sufficient fast rotation). 
Scaling laws for rotating finite amplitude convection have been
examined by Stellmach \& Hansen (2004).
They confirmed the above
denoted scaling law for the size of a convection cell by evaluating the
preferred wave number $k=1/\lambda^{\mathrm{corr}}$ from the
maximum value of the kinetic energy spectra. 
Within the rather restricted range of rotation rates our results reconfirm
this scaling for  $\lambda^{\mathrm{corr}}_x$:
\[
\frac{\lambda^{\mathrm{corr}}_x(\mathrm{Ta}=10^6)}{\lambda^{\mathrm{corr}}_x(\mathrm{Ta}=10^7)}\approx
1.5 \approx \left(\frac{10^6}{10^7}\right)^{-\frac{1}{6}}.
\]

The correlation length
parallel to the imposed field, $\lambda^{\mathrm{corr}}_y$, resembles the
increasing extension of the convection cells in direction of the imposed field. 
The field strength at which the transition to an anisotropic, elongated cell occurs
only weakly depends on the rotation rate, however, the increase of $\lambda_y^{\mathrm{corr}}$ is
slightly stronger for slower rotation.

For increasing co-latitude $\theta$ the convection cells become more and more
tilted (according to the Taylor-Proudman theorem).  For a higher co-latitude,
$\lambda_{x}^{\rm{corr}}$ increases due to this geometric effect, whereas
$\lambda_{y}^{\rm{corr}}$ remains nearly unaffected (corresponding plots are omitted).

The anisotropic character of the turbulence is also apparent in the behavior
of the turbulence intensities $\langle u'^2_i\rangle$. For a quantitative examination the following
functions are introduced:
\begin{equation}
A_{\mathrm{H}}=\frac{\langle u'^2_y\rangle -\langle
  u'^2_x\rangle}{u_{\mathrm{rms}}^2} \qquad A_{\mathrm{V}}=\frac{\langle
  u'^2_x\rangle+\langle u'^2_y\rangle-2\langle u'^2_z\rangle}{u_{\mathrm{rms}}^2}
\end{equation}
(see also K\"apyl\"a et al., 2004, and
note the different definition for $A_{\mathrm{V}}$).
The anisotropy of the turbulence intensities is described by 
$A_{\rm{H}}$ and $A_{\rm{V}}$ in the following way: 

\medskip

\begin{math}
A_{\rm{H}} \quad \left\{
\begin{array}{ccl}
<& 0 & \mbox{dominated by turbulence perpendicular to $B_y$}\\
=& 0 & \mbox{horizontal isotropy}\\
>& 0 & \mbox{dominated by turbulence parallel to $B_y$}\\
\end{array}
\right.
\end{math}

\medskip

\begin{math}
A_{\rm{V}} \quad \left\{
\begin{array}{ccl}
<& 0 & \mbox{dominated by vertical turbulence}\\
>& 0 & \mbox{dominated by horizontal turbulence}\\
\end{array}
\right.
\end{math}

\medskip

\begin{figure}
\includegraphics[width=8.5cm]{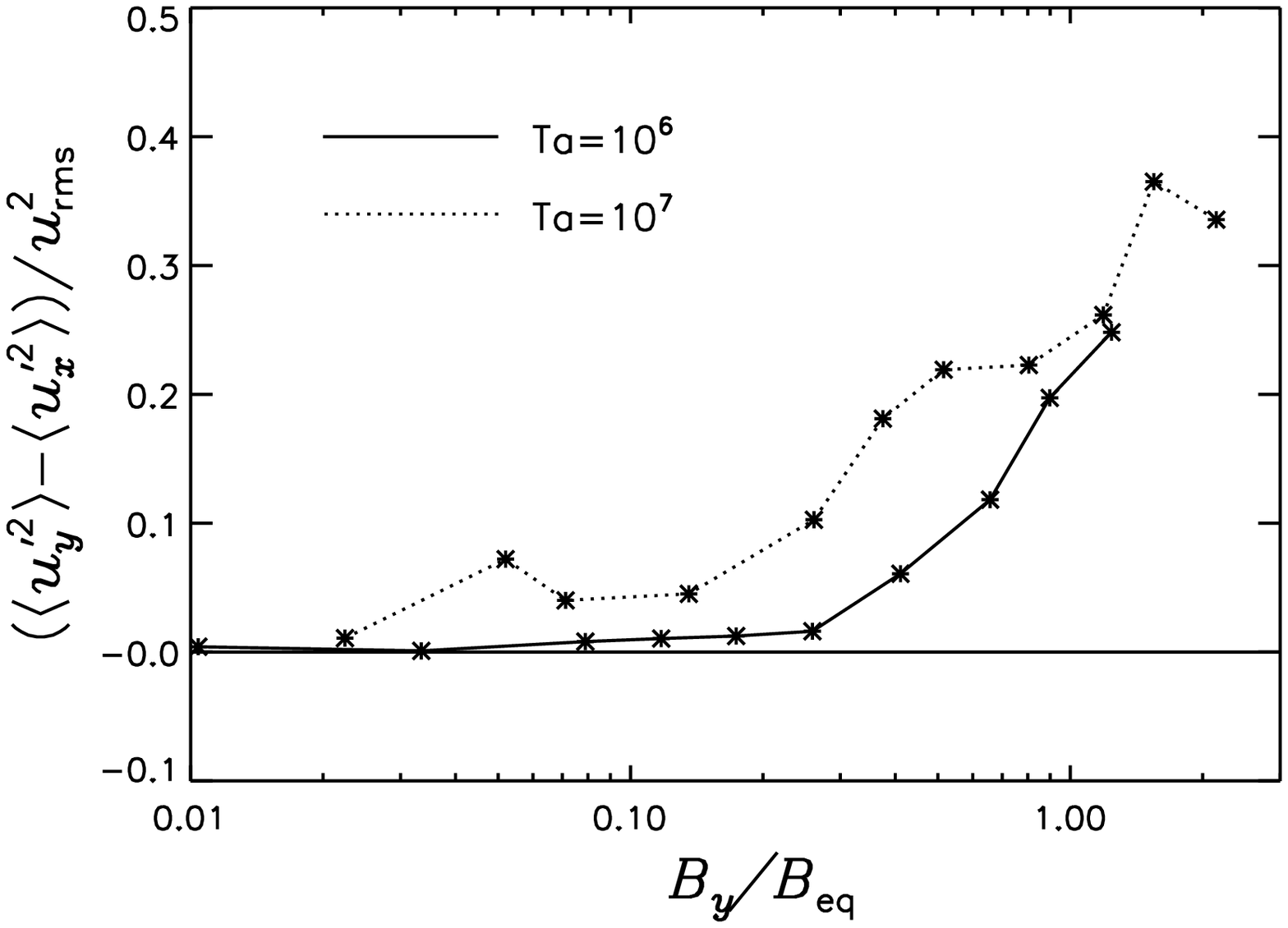}
\\
\includegraphics[width=8.5cm]{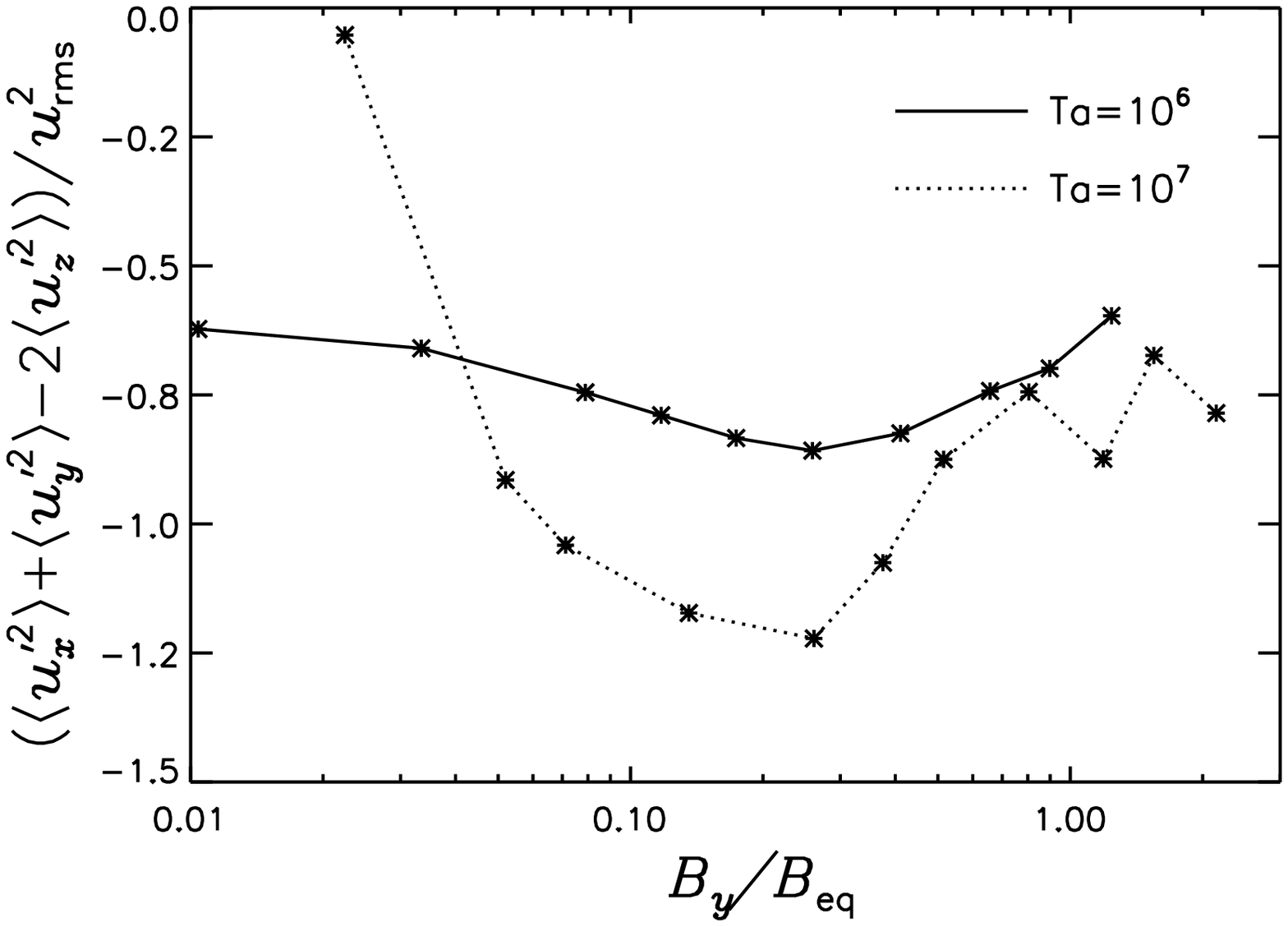}
\caption{Upper panel: Time and volume average of the horizontal anisotropy $A_{\mathrm{H}}$ for
  $\rm{Ta}=10^6$ (solid line) and  $\rm{Ta}=10^7$ (dotted line). Lower Panel:
  Time and volume average of the vertical anisotropy $A_{\mathrm{V}}$ for
  $\rm{Ta}=10^6$ (solid line) and  $\rm{Ta}=10^7$ (dotted line). $\theta=0^{\circ}$.\label{ratio}}
\end{figure}

As expected, in the presence of a magnetic field the motions perpendicular $B_y$ are
more strongly suppressed than the motions parallel $B_y$.
The resulting behavior of $A_{\mathrm{H}}$ at $\theta=0^{\circ}$ is shown in
the upper panel of Fig.~\ref{ratio}
which denotes the transition from the
horizontal isotropic turbulence to an anisotropic state. 
A dependence on the rotation rate is evident.
For $\mathrm{Ta}=10^6$ we approximately obtain
horizontal isotropy up to field strengths corresponding to $B\approx 0.4B_{\mathrm{eq}}$ ($\Lambda\approx
10$) whereas faster rotation results in $A_{\mathrm{H}} > 0$ except for very weak
magnetic fields.

The vertical anisotropy, $A_{\rm{V}}$, is presented in the lower panel of Fig.~\ref{ratio}.
For negligible magnetic fields an increase of the rotation velocity leads
to an isotropic behavior on the small scales ($A_{\mathrm{V}}\rightarrow 0$ for $B\rightarrow 0$), a property that has
been found by Brummell et al. (1996). 
Already weak horizontal fields break this tendency and result in a turbulence which is dominated by the vertical component.
Independent from the rotation rate the vertical anisotropy exhibits a
maximum around $B_y\approx 0.3B_{\mathrm{eq}}$.
The dominance of the vertical component is characteristic for the behavior of
the turbulence close to the pole and is not maintained for higher
co-latitudes.
The transition from vertical to horizontal dominated turbulence is determined by the orientation of the coherent cell
structures, i.e. by the orientation along the rotation axis and by the
elongation parallel to the dominant field component.
\subsection{Turbulent heat flux}
\subsubsection{General description of the turbulent transport of heat}

\begin{figure}
\includegraphics[width=4.3cm]{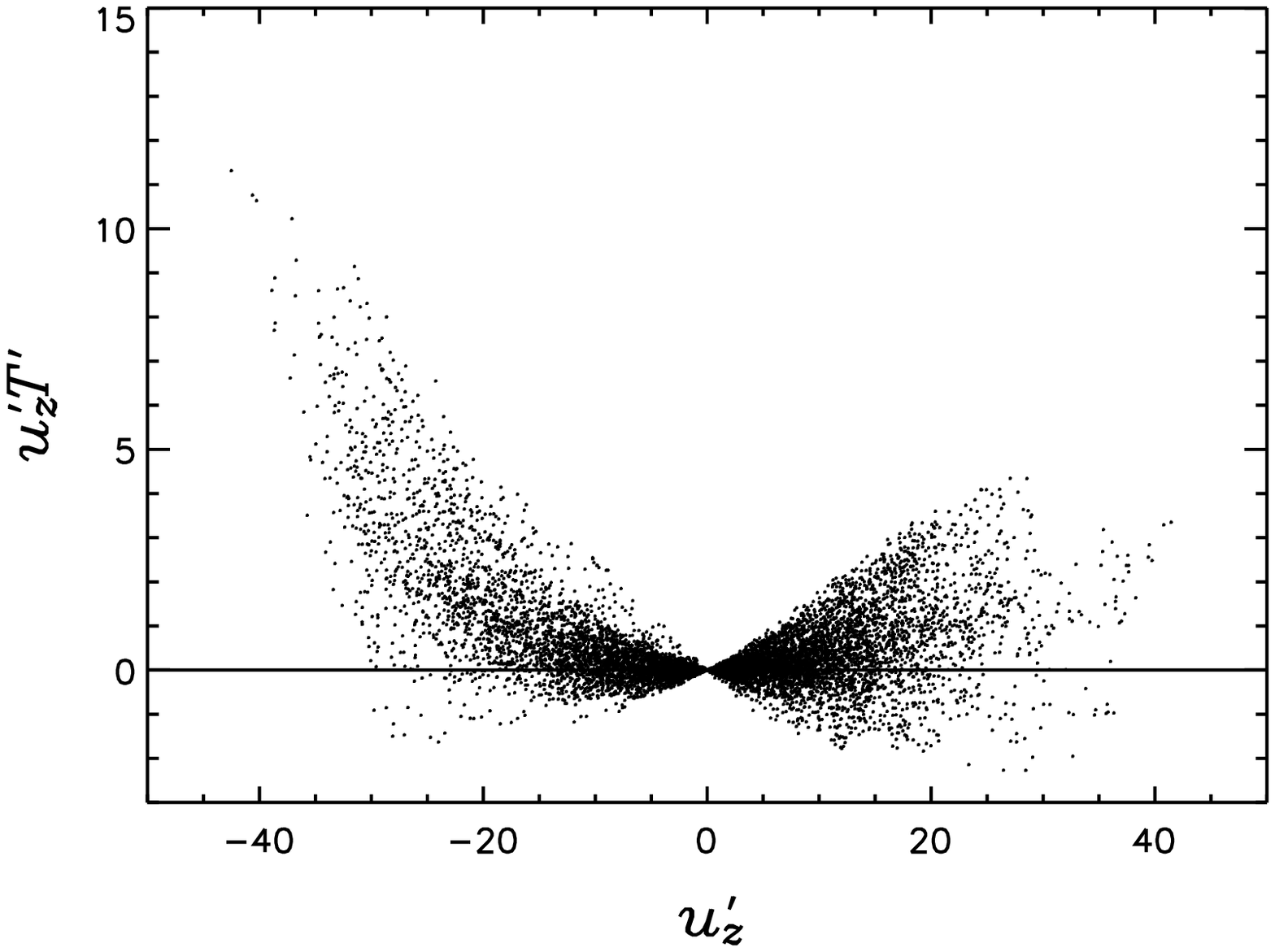}\nolinebreak[4!]
\includegraphics[width=4.3cm]{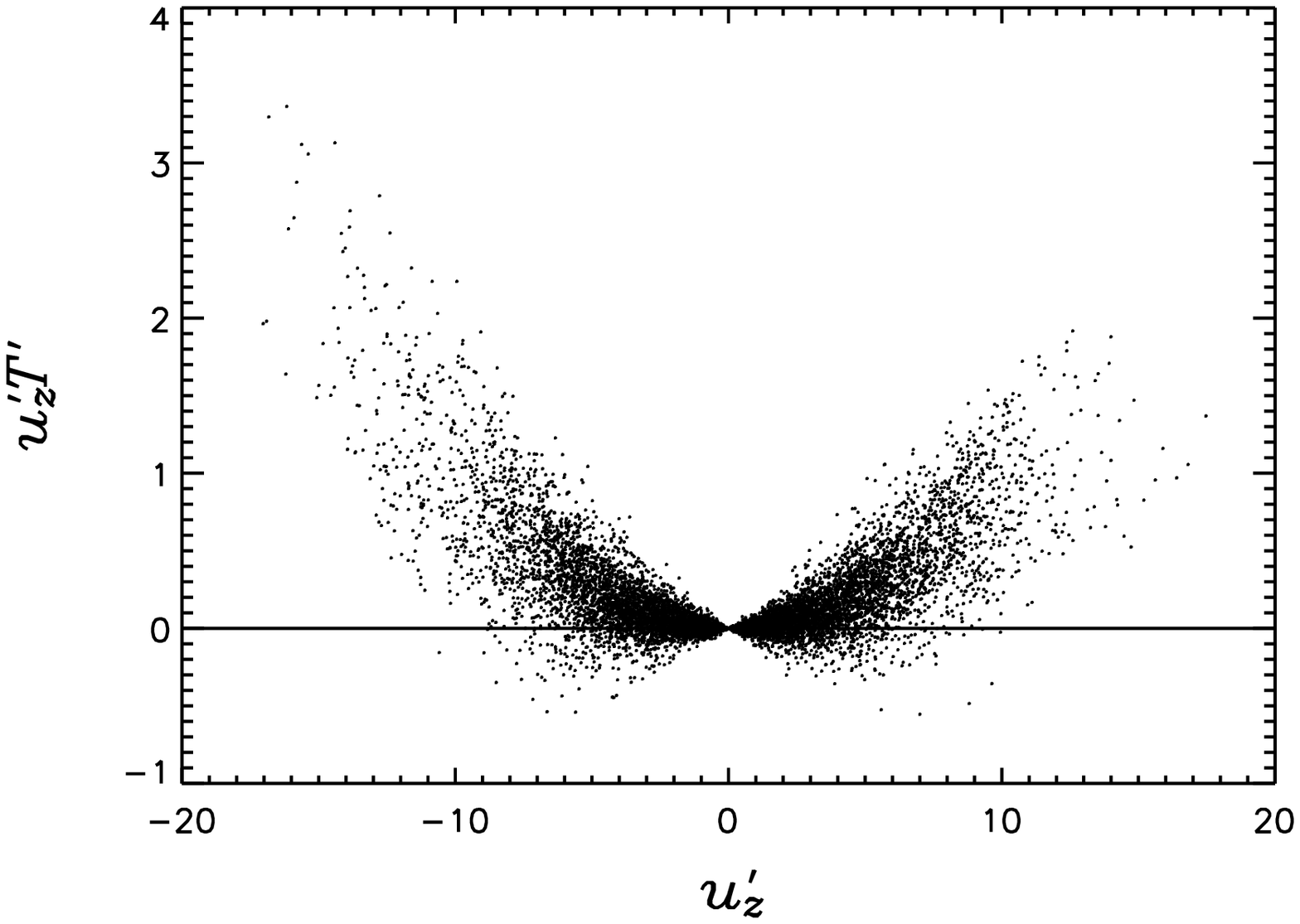}
\\
\includegraphics[width=4.3cm]{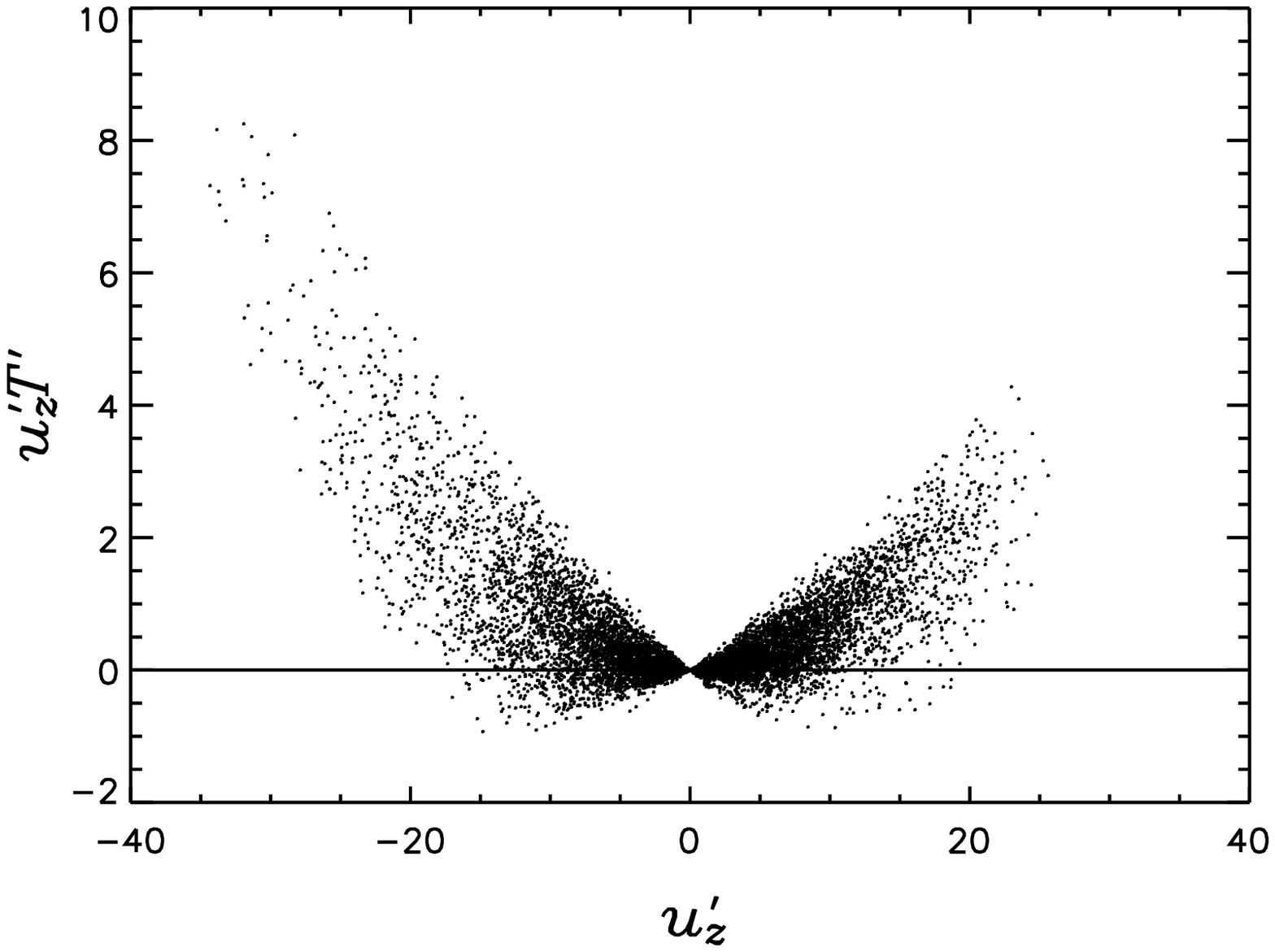}\nolinebreak[4!]
\includegraphics[width=4.3cm]{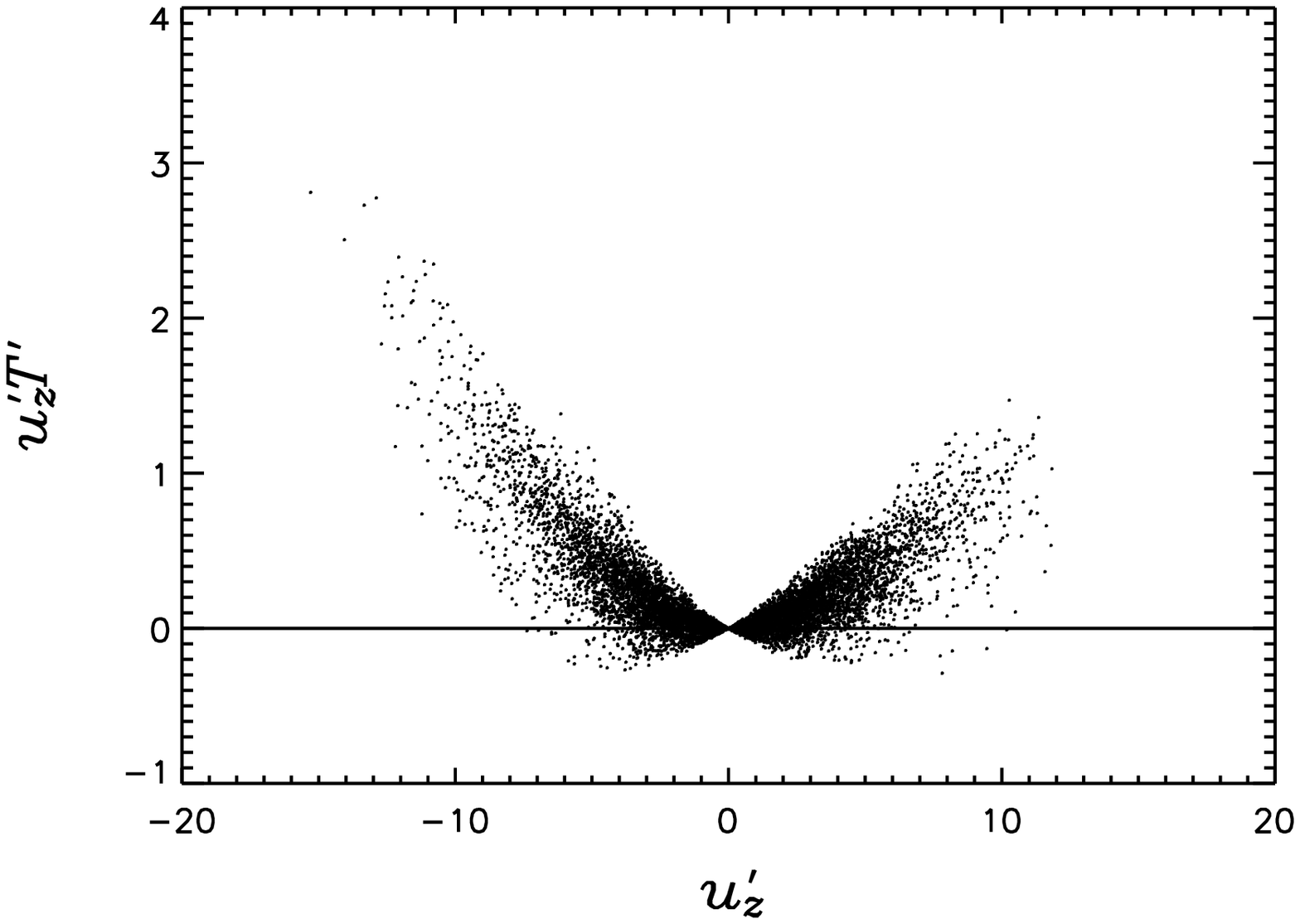}
\caption{Scatter plot of the normalized vertical heat flux in dependence of the turbulent
  up- and downflows at $z=0.5$. Upper left panel: convection, upper right panel: rotating
  convection (${\mathrm{Ta}}=10^7$, $\theta=45^{\circ}$) , lower left panel:
  magnetoconvection ($B_z\approx 0.2B_{\mathrm{eq}}$), lower right panel: rotating
  magnetoconvection, $\Lambda=1$ (corresponds to $B_y\approx 0.2B_{\mathrm{eq}}$), ${\mathrm{Ta}}=10^7$ and $\theta=45^{\circ}$.
}
\label{scatter}
\end{figure}

For statistically steady convection the turbulent heat flux
$F^{\mathrm{conv}}_i$ is given by: 
\begin{equation}
F^{\mathrm{conv}}_i={\rho c_p}\left<{u_i}'T'\right> =
  {\kappa_{ij}}\left(\frac{\partial T}{\partial x_j}-\frac{g_j}{c_{{p}}}\right)
\label{hf}
\end{equation}
(see e.g. R\"udiger, 1989) where $\kappa_{ij}$ denotes the tensorial heat conductivity  which is related to
the thermal diffusivity tensor $\chi_{ij}$ by $\kappa_{ij}=\rho c_{p}\chi_{ij}$.
In the presented model the large-scale temperature gradient $\partial
T/\partial x_j$ and the
gravity $g$
are constant and oriented parallel to the $z$-axis so that we are only able to discuss the
case $j=z$.
Since, in addition, the density $\rho$ remains approximately 
constant within the computational domain the simple relation 
$\kappa_{iz}\propto \langle u'_iT'\rangle$
holds to a good approximation and, in principal,
$\kappa_{iz}$ can immediately be determined from the numerically  computed correlations $\langle u'_iT'\rangle$.
It is convenient to define a normalized heat flux
\begin{equation}
{\widehat{F}}^{\mathrm{conv}}_i=\frac{F^{\mathrm{conv}}_i}{\rho c_p}=\langle
u'_iT'\rangle=\chi_{iz}\left(\frac{\partial T}{\partial z}
-\frac{g_z}{c_p}\right)
\end{equation}
which is the
quantity that is discussed with regard to the turbulent transport of heat in
the remainder of this paper.
The vertical heat flux is caused by
turbulent upflows of warmer fluid and downflows of cooler fluid. Mainly the fastest
up- and downflows contribute to the net flux which can be seen in 
Fig.~\ref{scatter} where time snapshots of the normalized flux $u'_zT'$ evaluated in a horizontal plane at
$z=0.5$ are presented in various scatter plots. 
Each dot denotes $u'_zT'$ at a certain grid cell in
dependence of the local vertical velocity $u'_z$. 
The sum of all values (divided by the number of data points) represents the
horizontal average of $\langle u'_zT'\rangle$ at $z=0.5$.
The four panels show the results from simulations of convection
(upper left panel), rotating convection (upper right panel), magnetoconvection
(lower left panel) and rotating magnetoconvection (lower right panel). 
Qualitatively the behavior is similar for all
four rather different states which only distinguish in the amplitude of
$u'_zT'$ and of the velocity $u'_z$. 
The downflows ($u'_z <
0$) and the upflows ($u'_z > 0$) contribute with nearly the same amount to the net
flux (although it seems that the downflows are slightly prevailing). 
This differs from the results for a strongly stratified layer presented by
Ziegler (2002) who obtained a preponderance of the contributions by
the downflows in non-rotating magnetoconvection which does not occur in sufficient fast rotating
magnetoconvection. 

Typical volume averaged values of the vertical heat flux for moderate field
strength and $\rm{Ta}=10^7$ are given by $\langle
u'_zT'\rangle\approx 0.2$ (in code units, see also section 3.2.2.). 
From Eq. (17) we therefore retrieve $\chi_{zz}\approx 0.4$,
which is about four times larger than the value for the molecular diffusivity
$\chi=0.1$ as it results from the input parameters (all numbers in code units,
see section 2.3 and 2.5). 
Although only moderate turbulence is examined, the obtained coefficient for
the vertical transport of heat exceeds the molecular one, so
that the transport of heat by turbulent motions is more effective than
the molecular transport through heat conduction.
\subsubsection{Vertical heat flux and turbulence intensity}
\begin{figure}
\includegraphics[width=8cm]{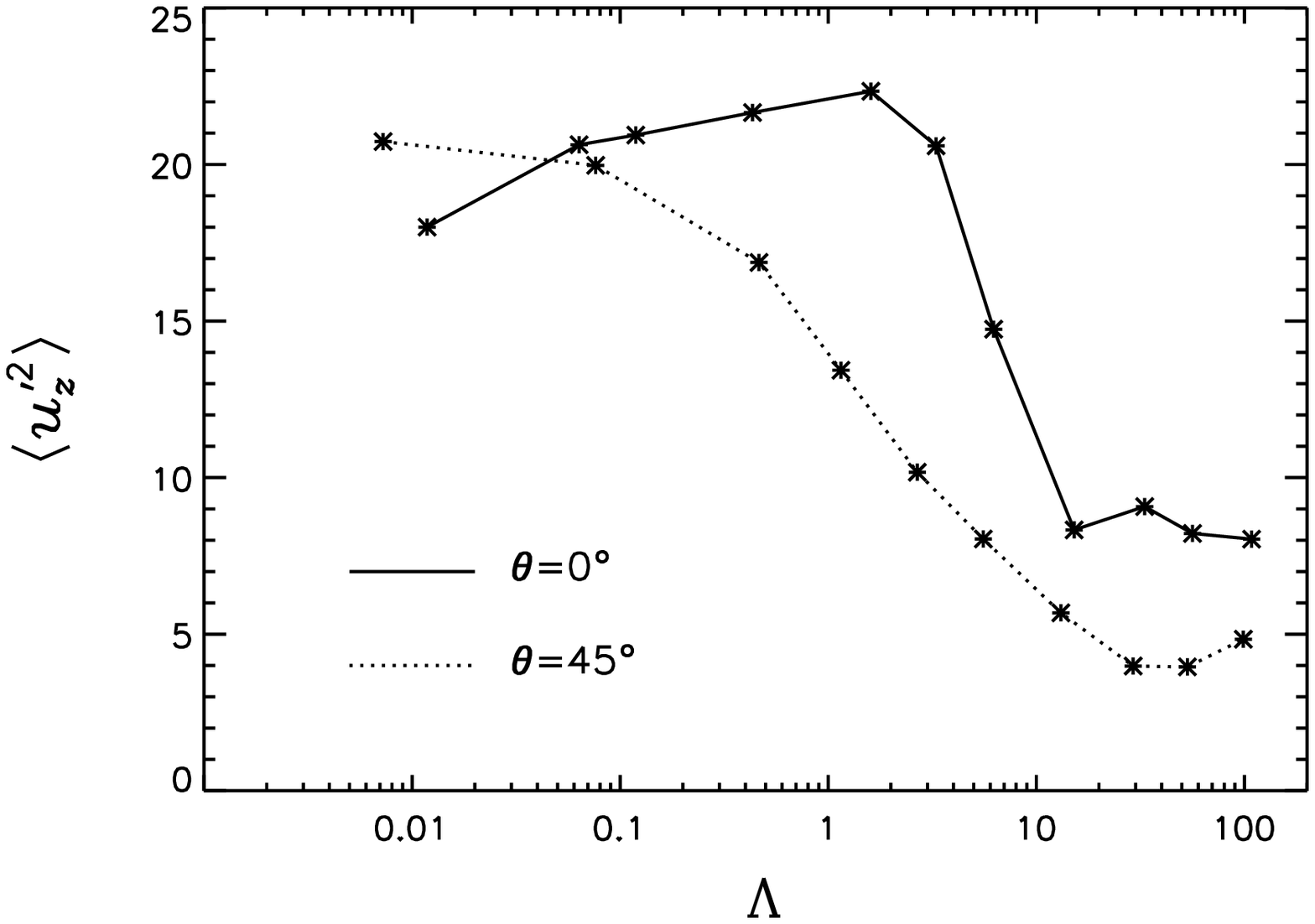}
\\
\includegraphics[width=8cm]{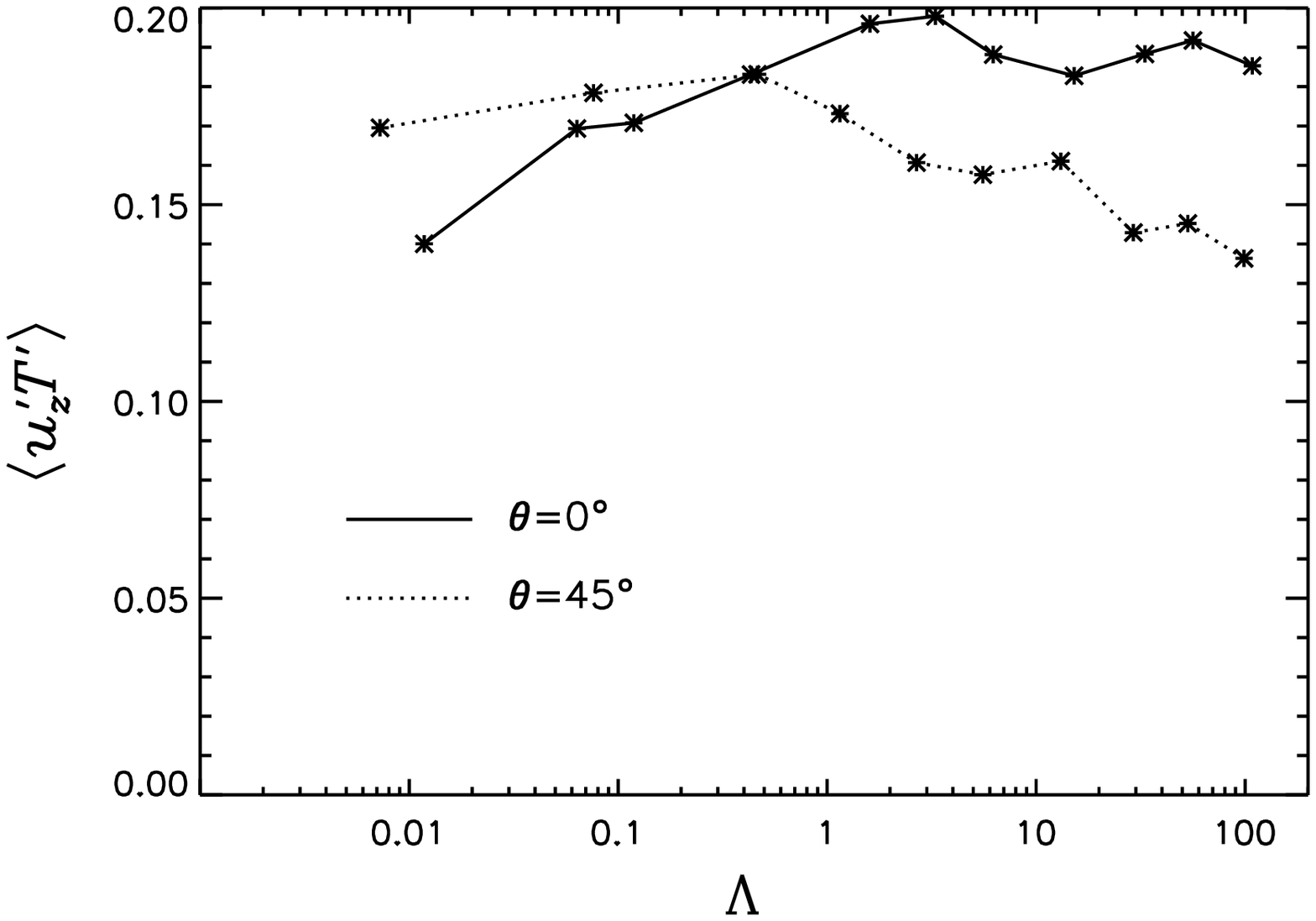}
\caption{Time and volume average of the turbulence intensity and heat flux in dependence of the Els\"asser
  number $\Lambda$. Upper panel: $\langle
  u'^2_{z}\rangle$, lower panel: $\langle
  u'_{z}T'\rangle$. $\mathrm{Ta}=10^7$.}
\label{hor_intensity__b}
\end{figure}

\begin{figure}
\includegraphics[width=8cm]{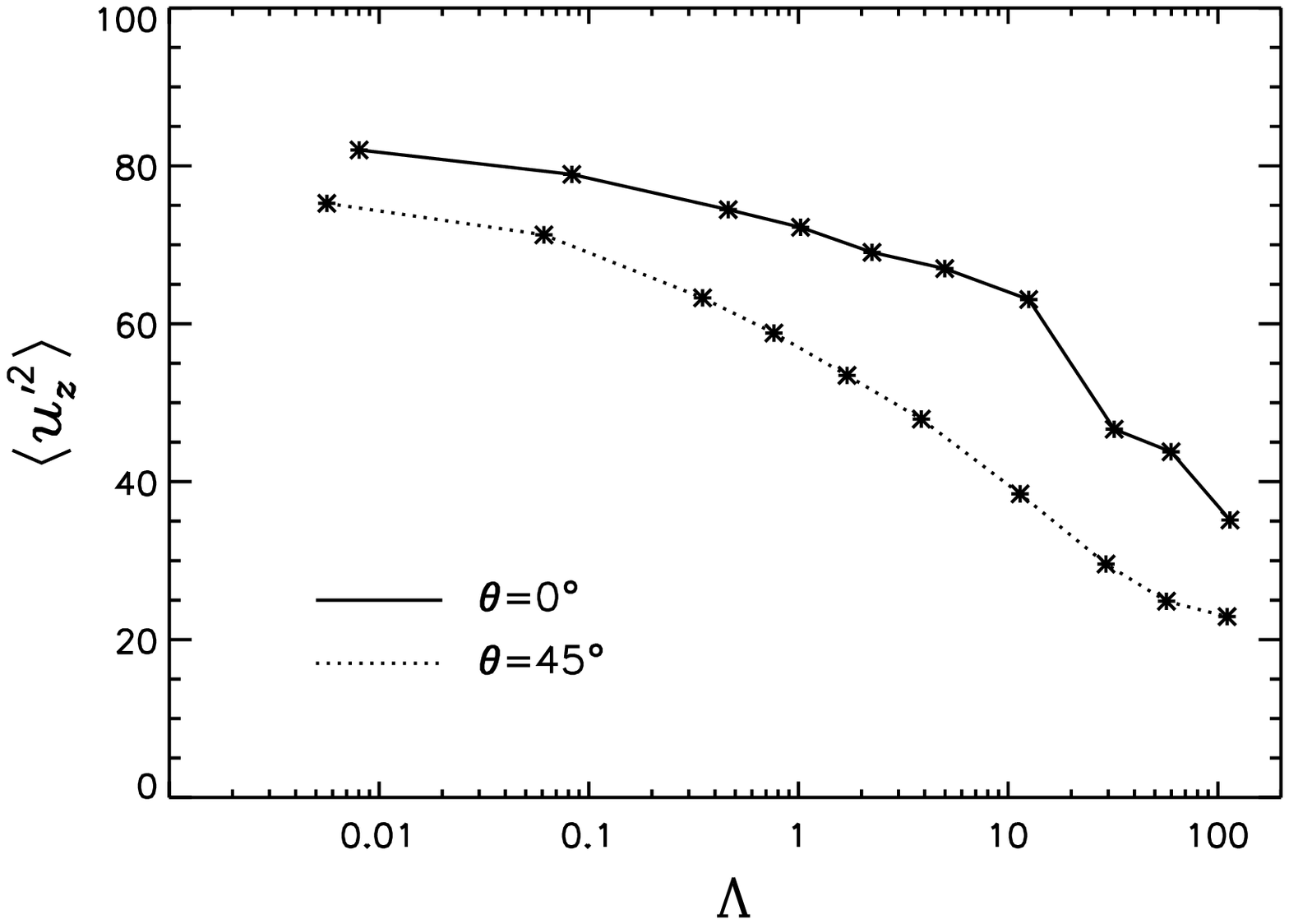}
\\
\includegraphics[width=8cm]{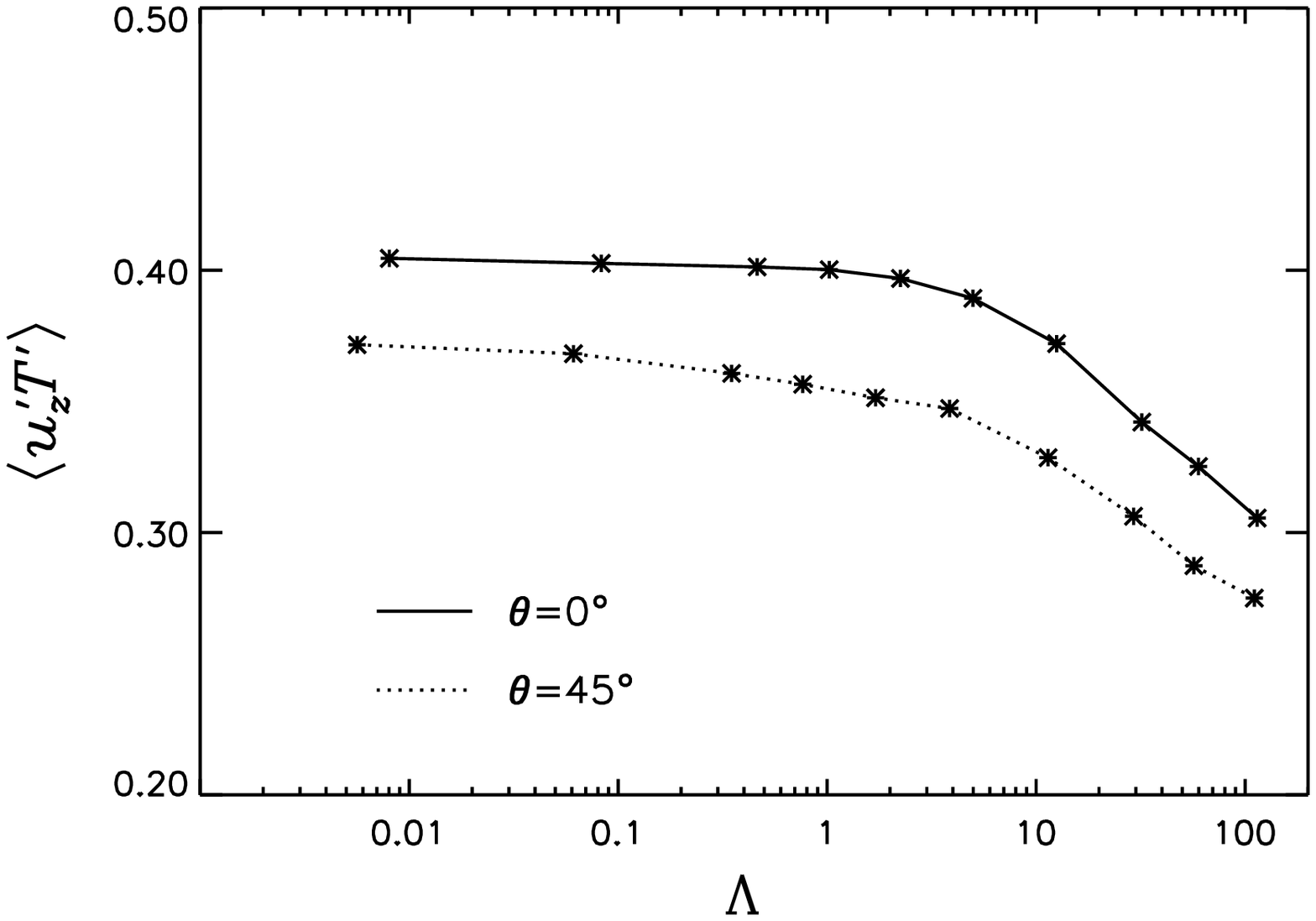}
\caption{Time and volume average of the turbulence intensity $\langle
  u'^2_{z}\rangle$ and the turbulent heat flux $\langle
  u'_{z}T'\rangle$ for $\rm{Ta}=10^6$.}
\label{hor_intensity__c}
\end{figure}

From the quasi-linear theory a simple expression relating the thermal
diffusivity tensor $\chi$ and $\widehat{\widetilde{Q}}_{ij}$, the Fourier
transform of the one-point correlation tensor $\widetilde{Q}_{ij}=\langle u'_i(\vec{x},t)u'_j(\vec{x},t)\rangle$ exists: 
\begin{equation}
\chi_{ij}=\int\!\int\frac{\chi k^2 \widehat{\widetilde{Q}}_{ij}(\vec{k},\omega)}{\omega^2+\chi^2k^4}d\vec{k}d\omega.
\end{equation} 
For $\chi\rightarrow 0$, the integrand can be replaced by a $\delta$-function, so that:
\begin{equation}
\chi_{ij}=\pi\int\widehat{\widetilde{Q}}_{ij}(k,0)dk = \frac{1}{2}\int
\widetilde{Q}_{ij}(0,\tau)d\tau\approx \frac{1}{2}\tau_{\rm{corr}}\widetilde{Q}_{ij}
\label{cond_usus}
\end{equation} 
where the $\tau$-integral is approximated by $\tau_{\rm{corr}}$, the correlation time of the turbulence.
A detailed explanation can be found in R\"udiger (1989).
This is a rather rough estimation which is based on the possibility to express
the temperature fluctuations $T'$ through the fluctuating velocity $u'$ and is valid in the mixing-length approximation, where the
turbulence spectra is dominated by one scale: $\sim \delta(k-\tau_{\rm{corr}}^{-1})\delta(\omega)$.
Eq.~(\ref{cond_usus}) might not be appropriate for a fast rotating system
under the simultaneous influence of a
magnetic field but it should deliver a general tendency for a relation between
the turbulent velocity fluctuations and the (turbulent) thermal diffusivity
tensor, respectively the normalized heat flux $\widehat{F}^{\mathrm{conf}}_i$.
In Fig.~\ref{hor_intensity__b} the behavior of the vertical turbulence
intensity $\langle u'^2_z\rangle$ (upper panel) is compared with the development of
the normalized vertical heat flux, $\langle u'_zT'\rangle$ (lower panel) for $\rm{Ta}=10^7$. 
Here, only the differences between the pole (solid line) and a co-latitude at
$\theta=45^{\circ}$ (dotted line) are pointed out. A more detailed analysis of the
angular dependence is given in the next section (\ref{secangular}).
As the most characteristic feature $\langle u'^2_z\rangle$ increases with increasing field strength and
exhibits a sharp drop for $\Lambda\ga 2$ at $\theta=0^{\circ}$. 
The maximum value for
$\langle u'^2_z\rangle$ is retrieved around $\Lambda\approx 1...2$. 
In this parameter regime (where
Lorentz force and Coriolis force are of the same order of magnitude) the
turbulence is enhanced compared to rotating non-magnetic convection.
This feature does not appear at $\theta=45^{\circ}$ or for a slower rotation
rate (Fig.~\ref{hor_intensity__c}).
The enhanced turbulence around $\Lambda\approx 1$ has been predicted by
Chandrasekhar (1961) who showed that the critical Rayleigh number
$\mathrm{Ra}_{\mathrm{crit}}$ is minimal if Lorentz and Coriolis forces are
comparable. 
In case of a fixed (overcritical) Rayleigh number a  drop of
$\mathrm{Ra}_{\mathrm{crit}}$ leads to a more overcritical convection which
results in a stronger driven flow and a more vigorous turbulence occurs.
Although the linear stability analysis of Chandrasekhar (1961) has been performed for a vertical oriented field
($B_z$), the general trend might also be true for a horizontal field. 
%
From the linear analysis it is also known that this effect is more
pronounced for faster rotating systems so that the maximum structure of $\langle u'^2_z\rangle$ will be more
dominant for even faster rotation, a result that already has been confirmed by Stellmach \& Hansen (2004).
However, the authors also obtained hints, that this effect -- predicted from a linear
stability analysis -- vanishes for stronger driven flows where the
non-linearities dominate the final state.

In accordance with the
behavior of $\langle u'^2_z\rangle$ the normalized vertical heat flux increases with the field strength below $\Lambda\approx
1$, but $\langle u'_zT'\rangle$ remains at a high level and only
weakly decreases for strong fields.
Although the fluctuating fluid motions are inhibited in the presence of a strong
magnetic field, this is obviously not the case for the temperature fluctuations, so that
the correlation of these two quantities is not suppressed with increasing field
strength.
This feature is not observable at a higher co-latitude where
$\langle u'_zT'\rangle$ exhibits a maximum around $\Lambda\approx 0.5$.

\begin{figure}
\includegraphics[width=8.5cm]{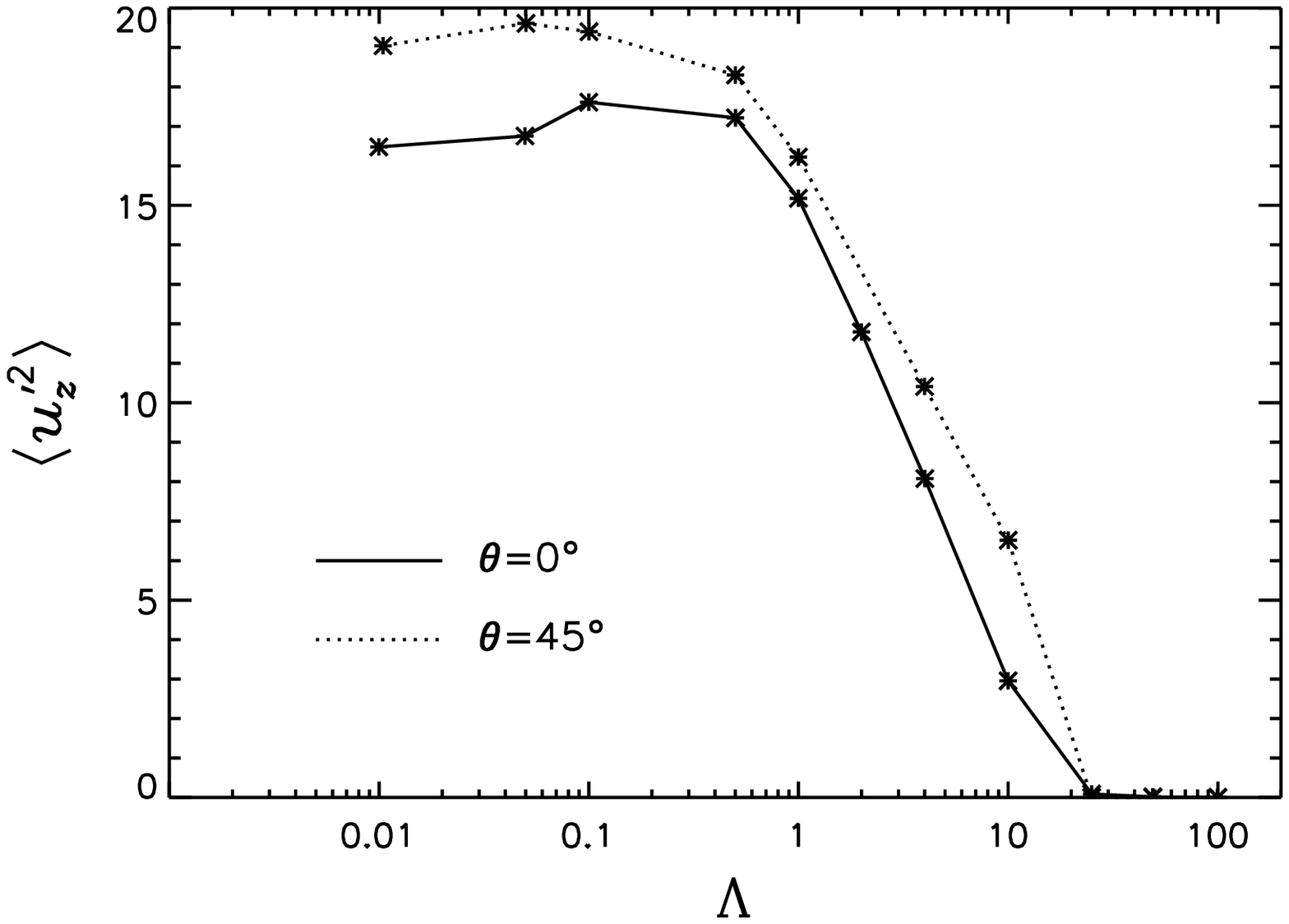}
\\
\includegraphics[width=8.5cm]{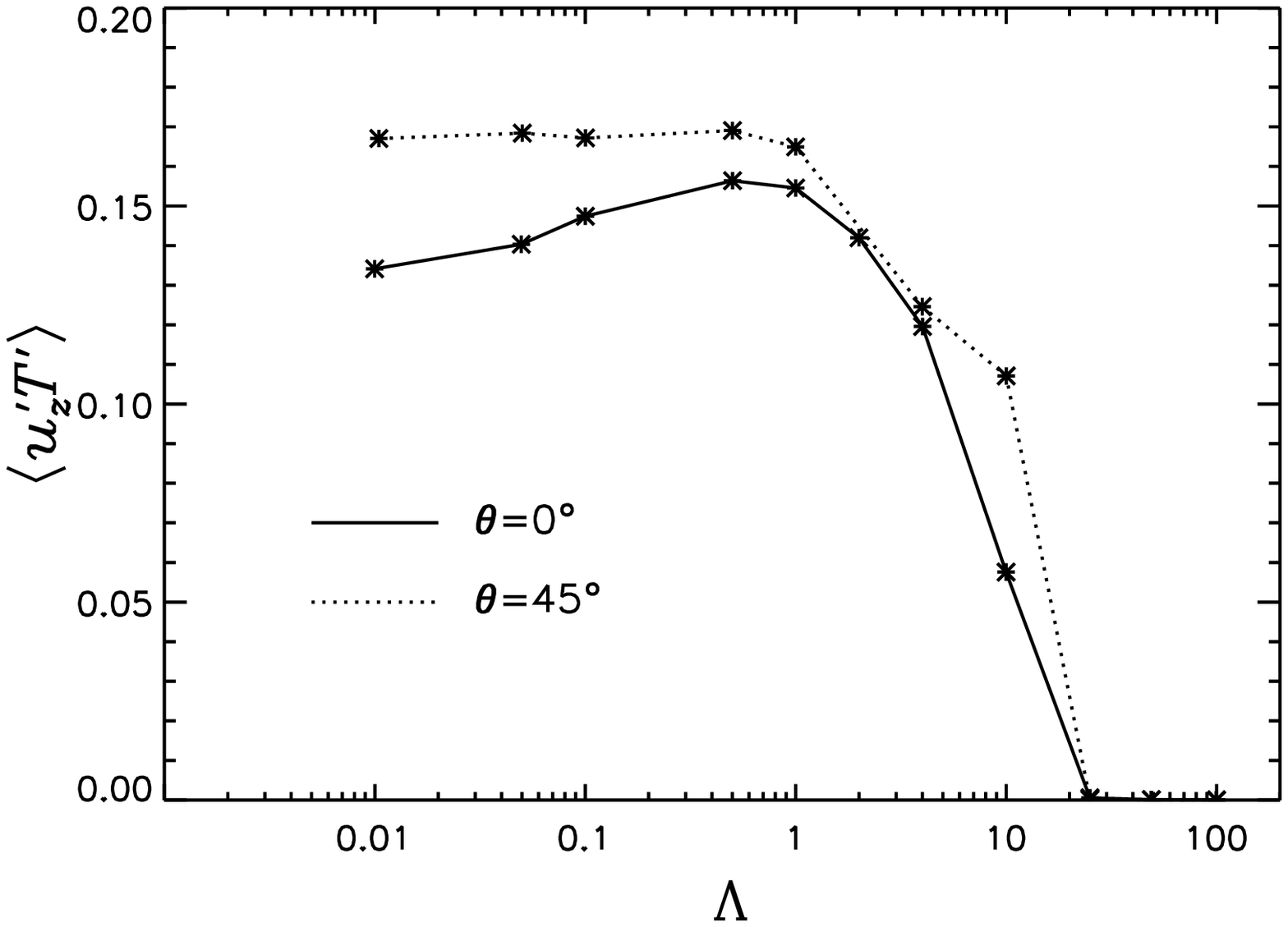}
\caption{Time and volume average of vertical turbulence intensity (upper panel) and vertical heat flux
  (lower panel) in case of a vertical oriented magnetic field ($B_z$). $\mathrm{Ta}=10^7$.}\label{vert_heatflux_vert_field}
\end{figure}

A second distinctive feature is reflected in the crossover of the two curves for
$\theta=0^{\circ}$ and $\theta=45^{\circ}$ in Fig.~\ref{hor_intensity__b}. 
For weak fields the turbulence
intensity and the vertical heat flux at $\theta=45^{\circ}$ are larger than at the
pole whereas both quantities behave the other way round for a field strength above
$\Lambda\approx 0.1...0.5$.
Such a behavior does not occur for slower rotation (see Fig.~\ref{hor_intensity__c}) or in
case of a vertically oriented field (Fig.~\ref{vert_heatflux_vert_field}).
The vertical turbulence intensity and heat flux under influence of
an external imposed $B_z$ are illustrated in Fig.~\ref{vert_heatflux_vert_field} (note, that in this case the magnetic boundary conditions
have to be changed to $B_x=B_y=0$ and $\partial_z B_z=0$).
Independent of the field strength, $\langle u'^2_z\rangle$ and
$\langle u'_zT'\rangle$ are slightly larger at $45^{\circ}$ and both quantities
show nearly the same dependence on the field strength.
At the pole both quantities exhibit a weakly pronounced maximum around
$\Lambda\approx 1$ and a very effective suppression occurs for $\Lambda\ga 2$.
\subsubsection{Horizontal heat fluxes and Reynolds stresses}
\begin{figure}
\includegraphics[width=8.5cm]{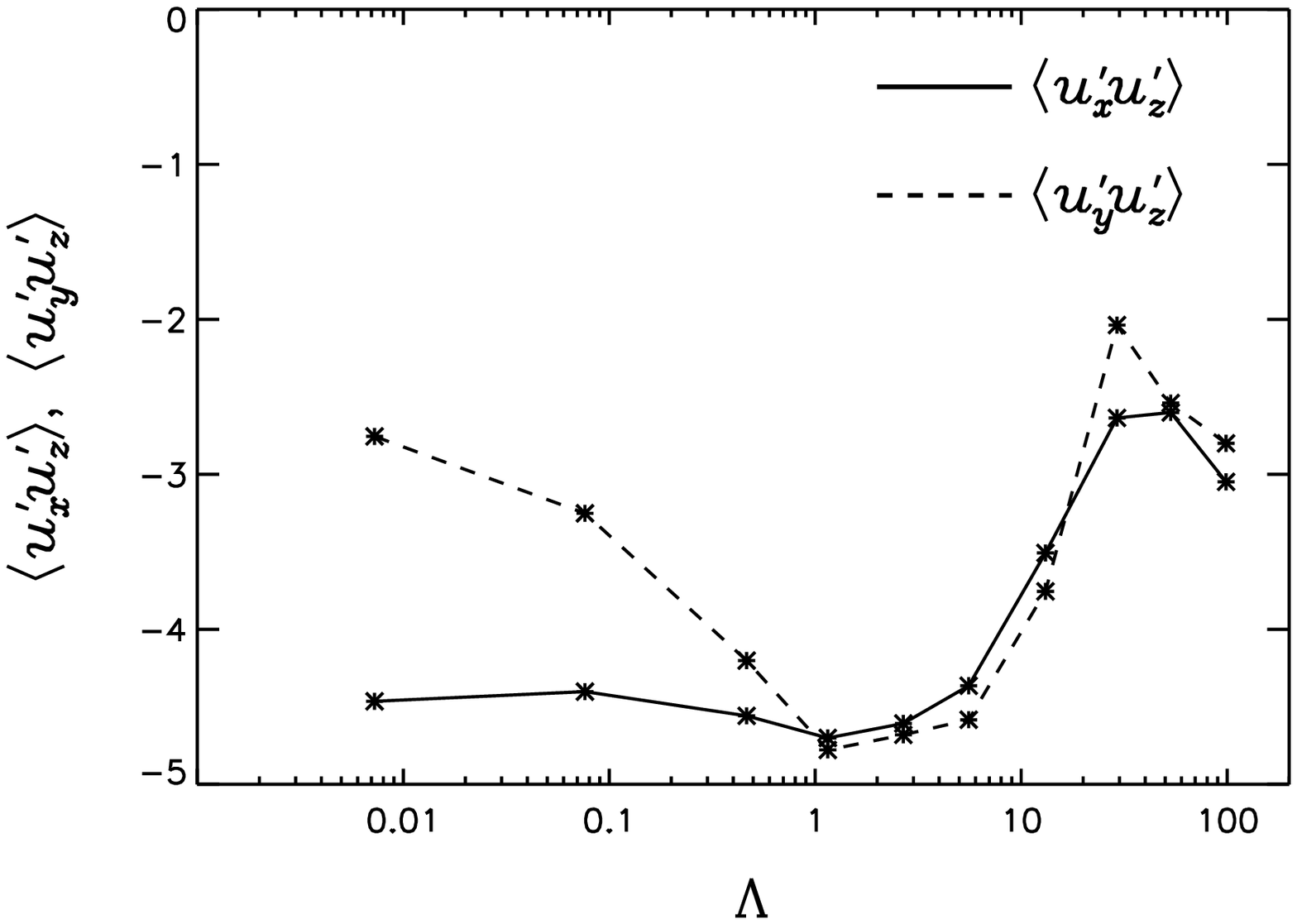}
\\
\includegraphics[width=8.5cm]{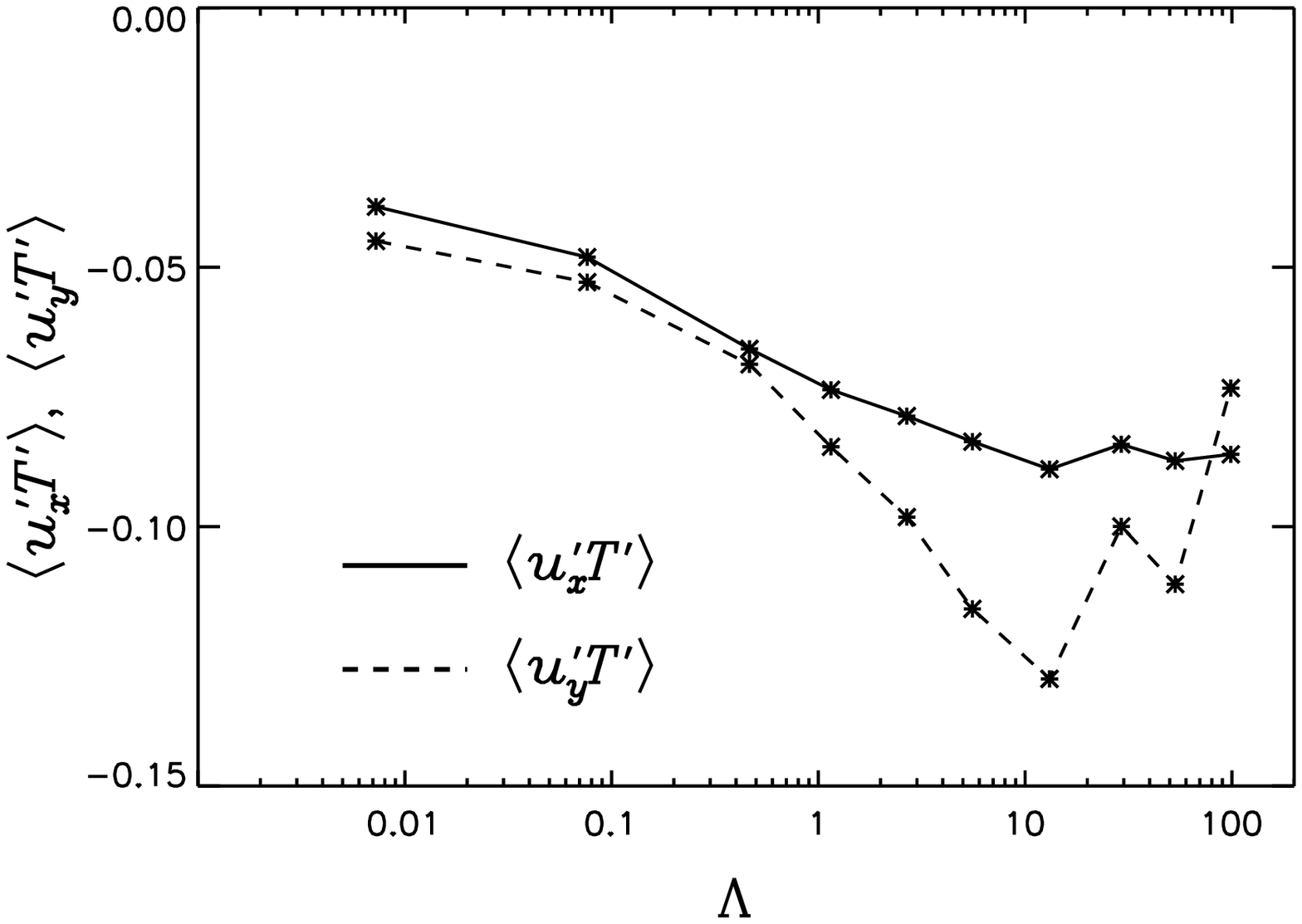}
\caption{Time and volume average of horizontal Reynolds stresses $\langle u'_{x,y}u'_z\rangle$ (upper panel) and
  normalized heat flux $\langle u'_{x,y}T'\rangle$ (lower panel) in dependence of the imposed magnetic field
  strength. $\theta=45^{\circ}$ and $\mathrm{Ta}=10^7$.}
\label{hor_flux__b}
\end{figure}

In the following, again, a horizontal field is imposed and the horizontal
transport of heat is discussed.
The horizontal components of the normalized heat flux, $\langle u'_{x}T'\rangle$ and $\langle u'_{y}T' \rangle$ and the corresponding
Reynolds stresses $\widetilde{Q}_{xz}$ and $\widetilde{Q}_{yz}$ vanish at the poles. This expected
behavior is nearly perfectly 
reflected in the numerical results and therefore the dependence on
the field strength is only
considered at $\theta=45^{\circ}$.
Figure~\ref{hor_flux__b} shows $\widetilde{Q}_{xz}$ and $\widetilde{Q}_{yz}$
(upper panel) in comparison with 
$\langle u'_xT'\rangle$ and $\langle u'_yT'\rangle$ (lower panel).
In all cases the considered quantities are
negative.
Therefore, the latitudinal component, $\langle u'_xT'\rangle$,
describes a transport of heat towards the (north-)pole.
The absolute value increases with increasing
field strength whereas the absolute value of $\widetilde{Q}_{xz}$ decreases with
increasing field strength.
As a consequence the relation between $\langle u'_xT'\rangle$ and $\widetilde{Q}_{xz}$
cannot be described by a simple expression given by Eq.~(\ref{cond_usus}).

In a spherical geometry the $y$-component of the heat flux corresponds to an azimuthal transport of
heat.
$\langle u'_yT'\rangle$, as it is
also always negative, describes a westward directed transport of heat.
The absolute value exhibits a distinct maximum value around
$\Lambda\approx 10$ and for very strong fields $\langle u'_yT'\rangle$
vanishes.
The behavior of the azimuthal heat flux roughly resembles the profile
of $\widetilde{Q}_{yz}$, although there exists a small offset for the location of the
minimum of $\widetilde{Q}_{yz}$ towards weaker magnetic fields.
\subsubsection{Angular dependence}\label{secangular}
In case of rotating turbulence without any other preferred
direction the turbulent diffusivity tensor $\chi_{ij}$ is given by 
\begin{equation}
\chi_{ij}=\chi_{\mathrm{T}}\delta_{ij}+\chi_{||}\frac{\Om_i\Om_j}{\Om^2}+{\widetilde{\chi}}\epsilon_{ipj}\Om_p
\label{cond_tensor_rot}
\end{equation}
(see e.g. Kitchatinov et al. 1994). 
$\chi_{||}$ describes enhanced thermal diffusion parallel to the rotation axis
and the expression ${\widetilde{\chi}}\epsilon_{ipj}\Om_p$ describes
enhanced/reduced diffusion in the azimuthal direction (which can be neglected
in an axisymmetric configuration). 
\begin{figure*}
\includegraphics[width=8.5cm]{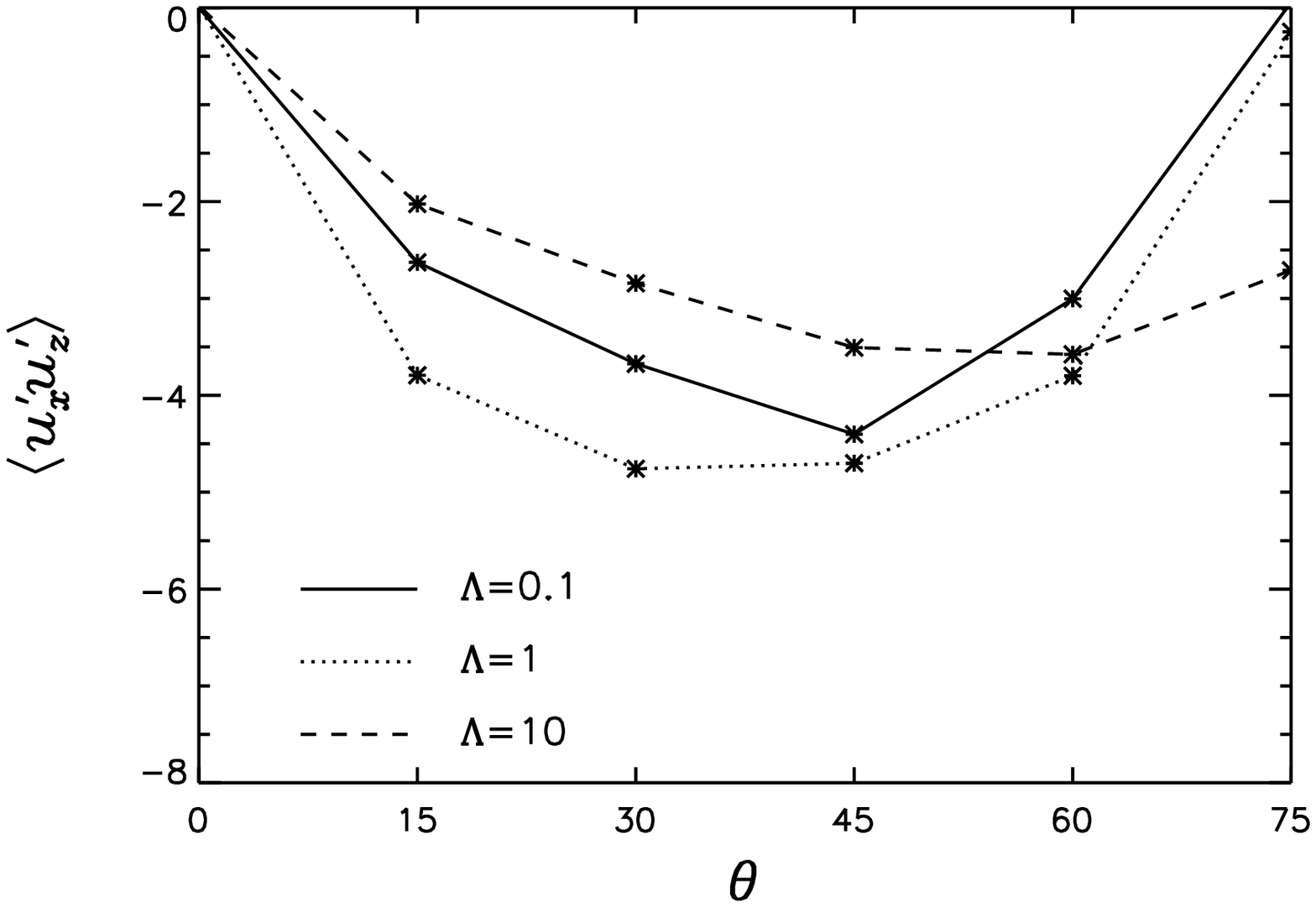}
\nolinebreak[4!]
\includegraphics[width=8.5cm]{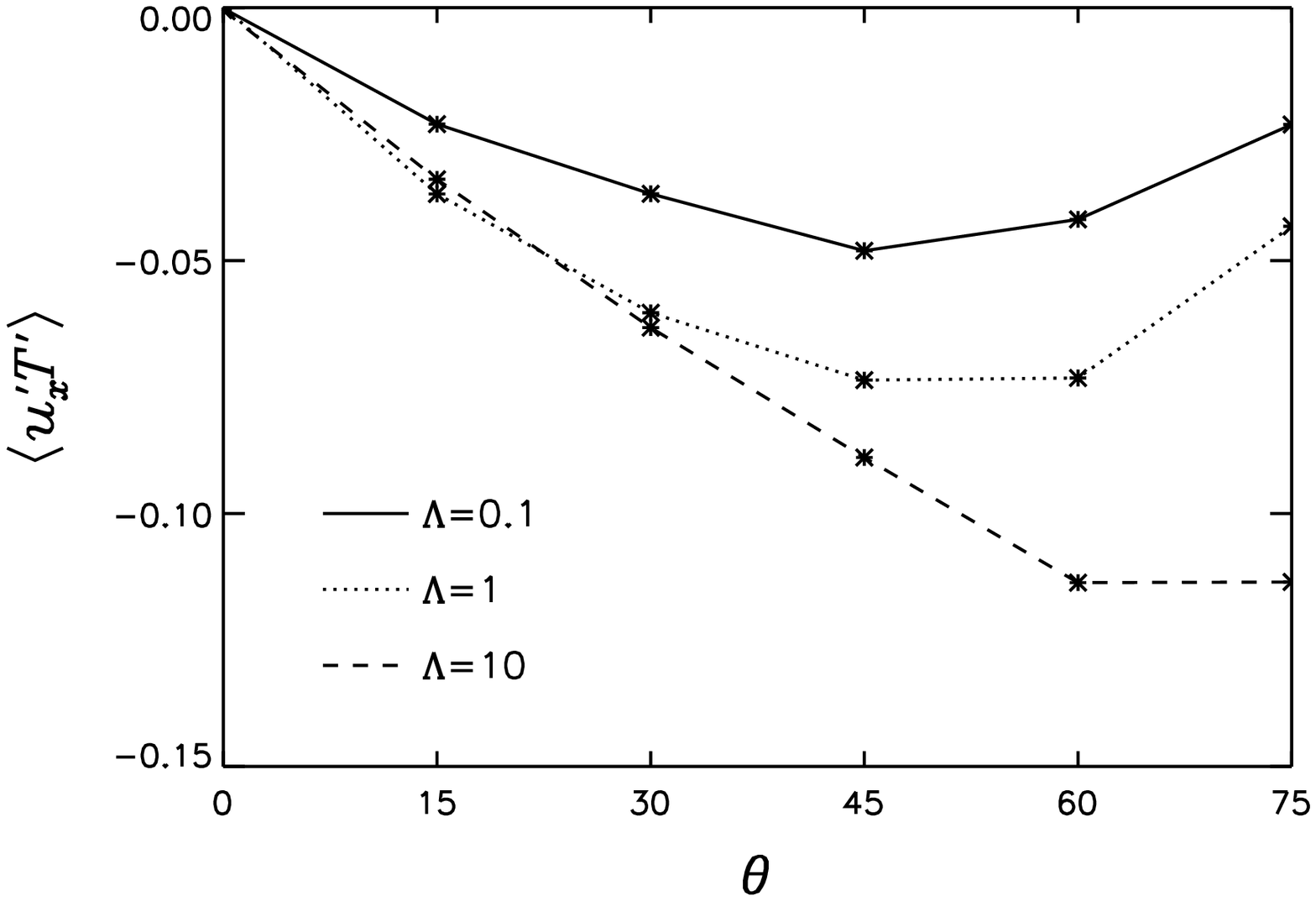}
\\
\includegraphics[width=8.5cm]{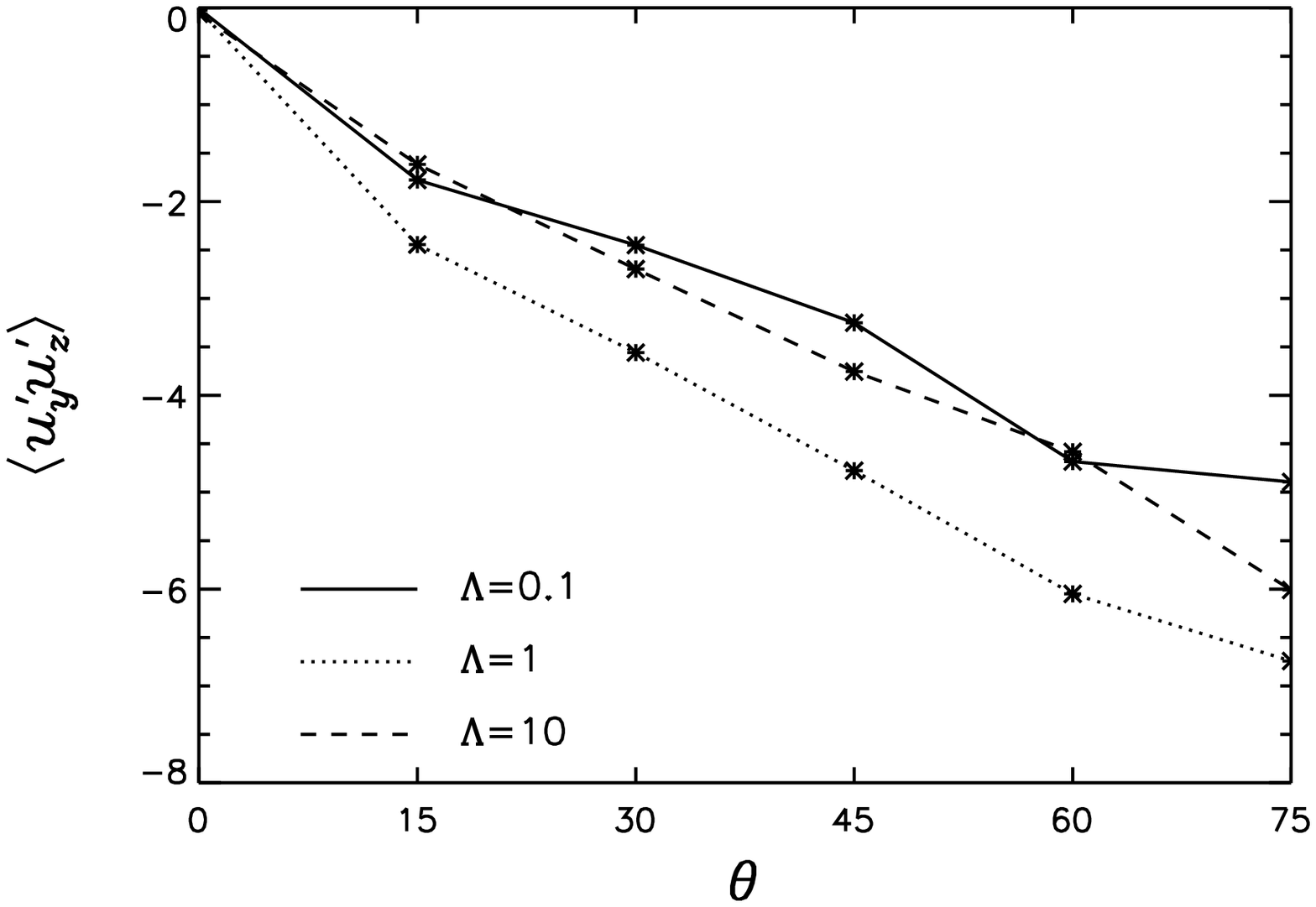}
\nolinebreak[4!]
\includegraphics[width=8.5cm]{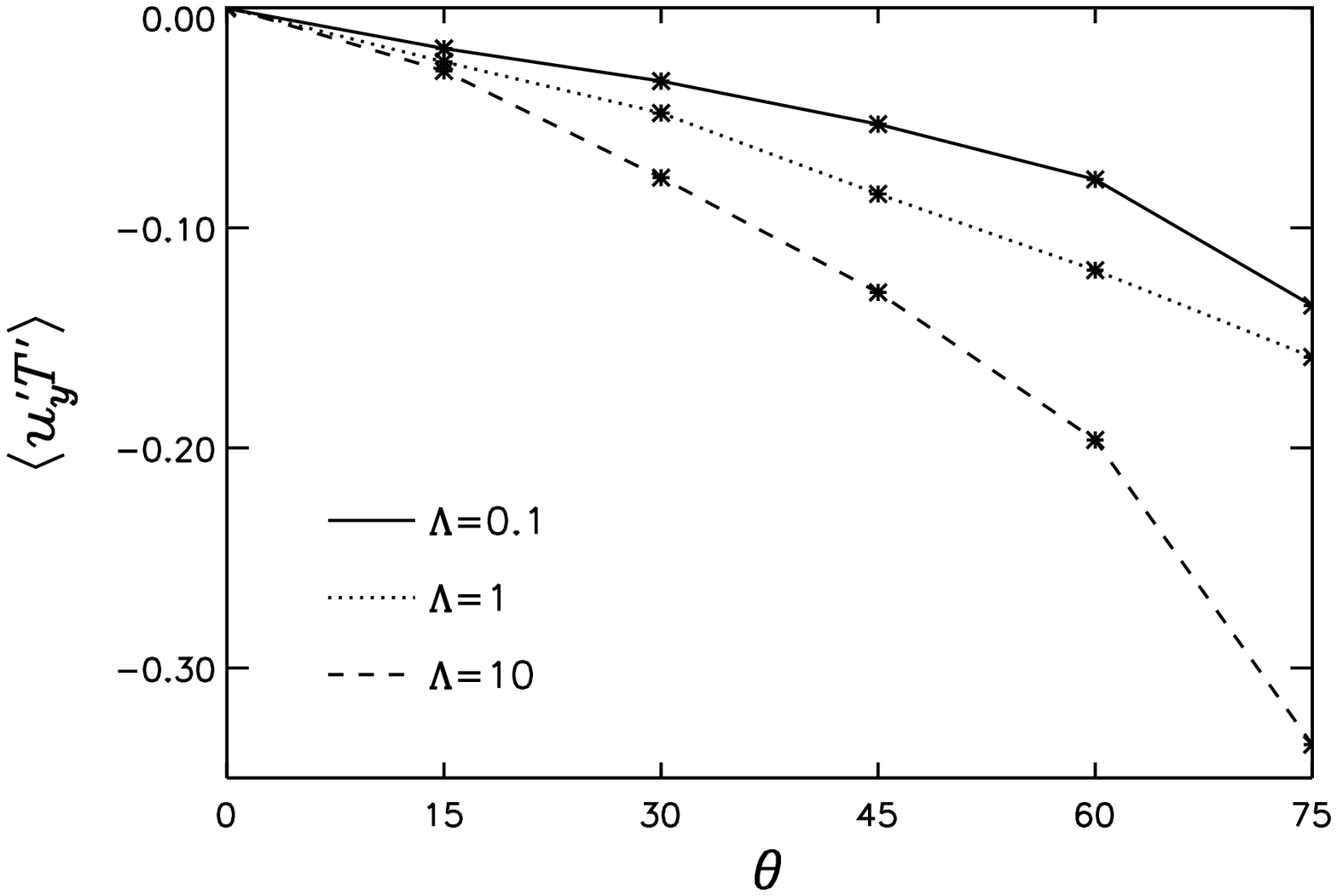}
\\
\includegraphics[width=8.5cm]{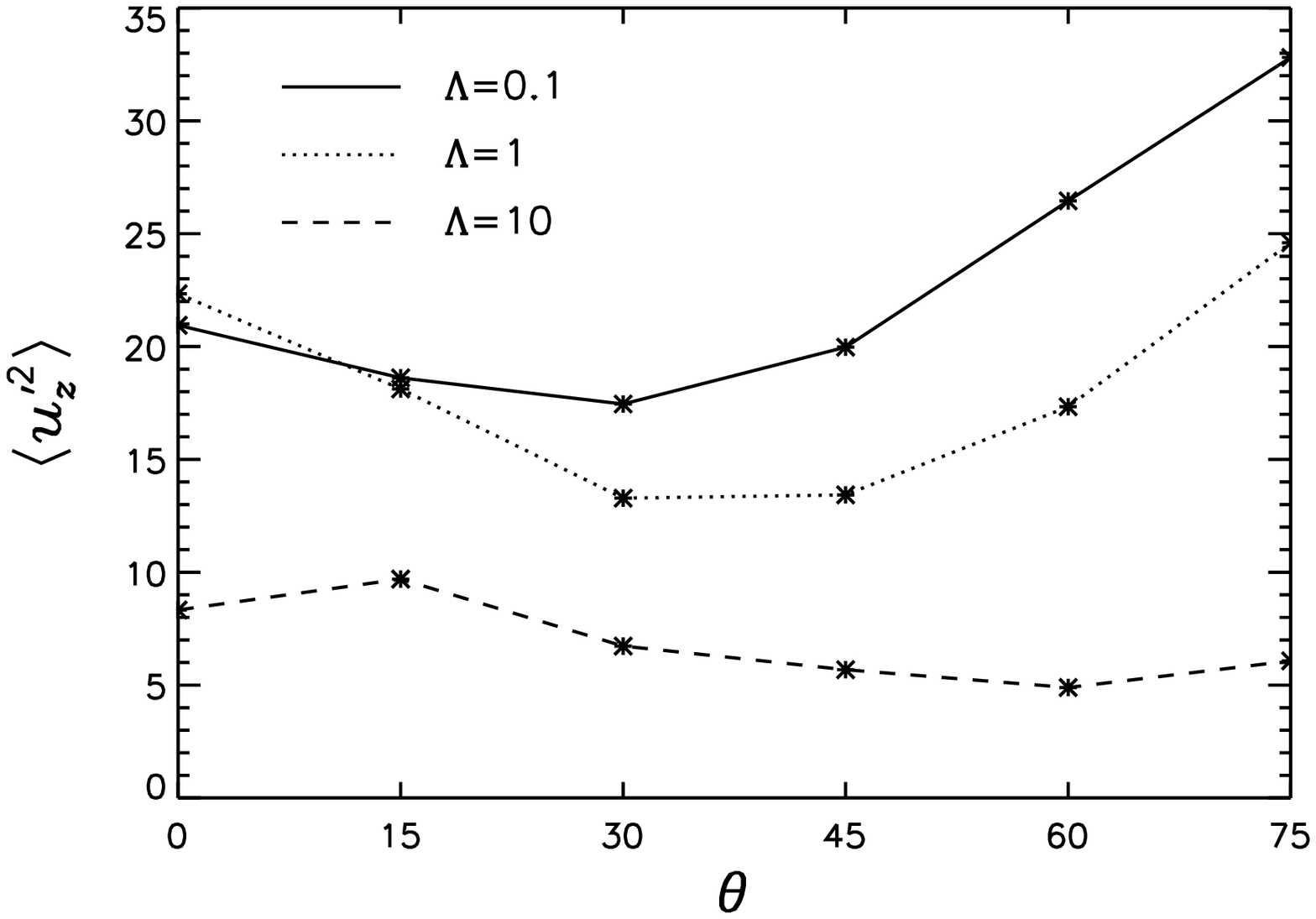}
\nolinebreak[4!]
\includegraphics[width=8.5cm]{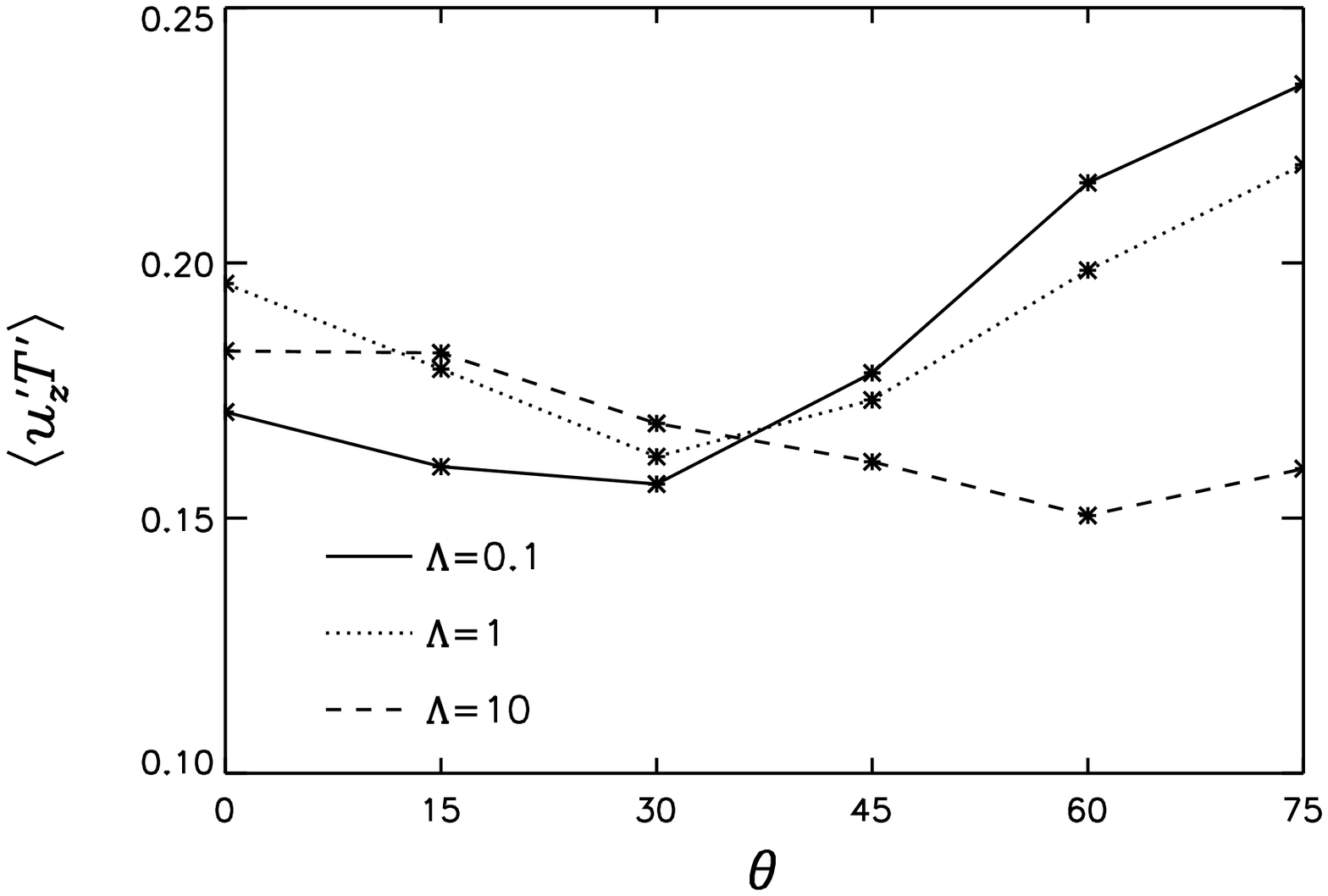}
\caption{Left side: Time and volume averages of the Reynolds stresses $\widetilde{Q}_{xz}, \widetilde{Q}_{yz}$ and
  $\widetilde{Q}_{zz}$. Right side: Time and volume averages of the heat fluxes $\langle u'_xT'\rangle$, $\langle u'_yT'\rangle$ and
  $\langle u'_zT'\rangle$. Both quantities are shown in dependence of the co-latitude $\theta$ for ${\mathrm{Ta}}=10^7$. The solid
  (dotted, dashed) curve denotes $\Lambda=0.1 (1, 10)$.}
\label{flux_lam_1_theta}
\end{figure*}
In the applied Cartesian system where we are restricted to a vertical large
scale temperature gradient the following three elements of $\chi_{ij}$ can be computed:
\begin{eqnarray}
\chi_{xz}&=&-\chi_{||}\sin\theta\cos\theta\label{cond_1}\\
\chi_{yz}&=&-\widetilde{\chi}\Om\sin\theta\label{cond_2}\\
\chi_{zz}&=&\chi_{\mathrm{T}}+\chi_{||}\cos^2\theta\label{cond_3}
\end{eqnarray}
These expressions neglect the influence of the magnetic field and might
therefore be valid only for small values of $\Lambda$.

The angular dependence of the Reynolds stresses and the normalized turbulent heat flux
are shown in Fig.~\ref{flux_lam_1_theta} (left hand side: $\widetilde{Q}_{iz}$; right hand
side $\langle u'_iT'\rangle$). 
The solid (dotted, dashed) lines denote the results for
$\Lambda=0.1 (1, 10)$ (corresponding to $B_y=0.07 (0.2, 0.7)B_{\mathrm{eq}}$).
In all cases the meridional component $\langle u_x'T'\rangle$
is negative. 
For $\Lambda \approx 0.1$ and $\Lambda\approx 1$ the
absolute value exhibits a maximum around $\theta=45^{\circ}$ and in both cases
$\langle u'_xT'\rangle$ can be described by a function
$\propto\sin\theta\cos\theta$ as indicated by equation (\ref{cond_1}). 

The azimuthal component $\langle u'_yT'\rangle$ is also always negative and
the absolute values increase towards the equator so that at least the tendency
coincides with the behavior indicated by expression
(\ref{cond_2}). 
$|\langle u'_yT'\rangle|$ increases with increasing $B_y$ which becomes more pronounced towards the equator.

The vertical component $\langle u'_zT'\rangle$  is always positive thus heat is transported outwards. 
For
$\Lambda\la 1$ the vertical heat flux increases towards the
equator and, in principle, the behavior could be
approximated by an expression of the form given by equation (\ref{cond_3}). 
This is only possible if $\chi_{||}$ is negative.
However, a negative value for $\chi_{||}$ is in contradiction with the results for the meridional heat flux
which require $\chi_{||} > 0$.
A positive value for $\chi_{||}$ is also obtained from calculations of
Kitchatinov et al. (1994).
Regarding the Coriolis force at the pole,
$\vec{\Om}\times\vec{u}=\Om_0(-u_y,u_x,0)$, it is evident that no (linear) coupling
between vertical and horizontal components is possible.
A transfer of vertical and horizontal momentum (through the coupling of $u_z$
with $u_x$ respectively $u_y$) is only possible at higher co-latitude.
Therefore one could suppose that quantities, like the vertical heat flux or the
vertical component of the turbulence intensity are maximum at the pole and
decrease towards the equator.
The somehow surprising
opposite behavior shown in the simulations here, is confirmed the numerical results of
K\"apyl\"a et al. (2004) and R\"udiger et al. (2005). 
K\"apyl\"a et al. (2004) calculated the vertical and the meridional heat flux
for rotating convection in a strongly stratified layer for a various number of
rotation rates. 
In all cases
they report the maximum of the vertical heat flux at the equator and ``a
tendency'' for the minimum at a co-latitude $\theta\approx
30^{\circ}...45^{\circ}$.
These results are confirmed by our calculations up to $\Lambda\approx 1$.
Only for strong magnetic fields ($\Lambda\approx 10$) the vertical turbulence intensity
and heat flux are larger at the pole.
\section{Discussion and Conclusions}
Rotating magnetoconvection has been examined for a wide range of
parameters and due to the varieties of the results it is nearly impossible to
describe the outcome uniformly. 
The convectively driven flow changes its characteristic properties considerably in dependence of
the applied magnetic field and/or rotation rate. 
The behavior of the cell pattern indicates that in case of a dominant toroidal field 
convection in a fast rotating system should
occur in extremely thin sheet like
structures. 
In the presented simulations the vertical extension of the cells 
is fixed by the vertical boundary conditions so that the average size of the
convection cells in $z$-direction is extended over the whole box domain.
It is hardly
believable that such structures are stable over the largest possible scales
given by the size of the fluid outer core  so that the ``real'' vertical
structure of the convection remains dubious and might
only be obtained from, at present unrealistic, global high-resolution simulations.
For faster rotation the effects of the magnetic field on the fluid flow
via the Lorentz force become more significant. 
This has been  shown by the behavior of the
anisotropy functions $A_{\mathrm{H}}$ and $A_{\mathrm{V}}$.
Particular attention deserves the vertical anisotropy. 
The isotropic behavior on the small scales caused by rotation vanishes with the
introduction of a magnetic field. 
As a consequence, the vertical component of the turbulence intensity dominates
the flow which leads to an enhanced vertical mixing of the convective layer. 
Turbulent transport in a rotating system is facilitated through the
introduction of a magnetic field.

The turbulent heat flux has been calculated as an example for a diffusive
process that occurs on scales that are not resolved in global simulations. 
Most remarkably, the vertical heat flux at the pole slightly increases with increasing
magnetic field strength (if the rotation is fast enough) and remains at a high
level even for very strong magnetic fields.
Thus, cooling of the Earth's interior is enhanced by the
presence of a dominant toroidal field.
Comparison of the results from non-rotating magnetoconvection with simulations that involve rotation and a horizontal (Fig.~\ref{hor_intensity__b}) 
or a vertically oriented magnetic field (Fig.~\ref{vert_heatflux_vert_field}),
respectively, shows that an enhancement of
the vertical heat flux at high field strength only occurs in combination of a
horizontally oriented magnetic field, a sufficient fast rotation
rate and close to the north-pole.
However, the effects of $B_y$ on the vertical heat transport are relatively
small compared to the strong quenching for $\Lambda\ga 2 (B\ga
0.3B_{\mathrm{eq}})$ in case of an imposed $B_z$.

Likewise of interest is the behavior of the meridional heat flux which is always
negative, indicating that heat is transported towards
the poles.
From comparison with slower rotating magnetoconvection we know that the
rotational quenching of $\langle u'_xT'\rangle$ is weaker than for
$\langle u'_yT'\rangle$ and $\langle u'_zT'\rangle$, so that it
might be possible that in a fast rotating system a significant amount of heat is transported towards the
poles.
Therfore, the poles should be warmer than the equator. 
This results in a non-radial
oriented large-scale temperature gradient which acts as a source for a
meridional large-scale flow.
Such a meridional flow -- if strong enough -- interacts with the column like
large-scale structure of the convective motions and might be important
for the understanding of the dynamo mechanism in the Earth's core (see e.g.
Sarson \& Jones (1999), who identified a fluctuating meridional flow as an
explanation for the occurrence of dipole reversals). 
Indeed, global simulations of Olson \& Christensen (2002) yield a warmer
northpole (see their Fig.~5) but it is not clear if this effect is a result of
turbulent heat transport.
A large scale meridional flow is also found in the solar convection zone,
giving rise to important consequences for the solar rotation law (R\"udiger
et al., 2005).

The distribution of the temperature on the top of the fluid core depends
on the heat flow at the core mantle
boundary that is usually prescribed as a boundary condition. 
Seismological measurements indicate a non-uniform heat flow at the
core mantle boundary, whose influence has been examined in numerical
simulations of Olson \& Christensen (2002). 
The lateral and longitudinal variations of the boundary heat flux affect the
local behavior of the fluid flow:
convection is enhanced (inhibited) where the boundary heat flow is high
(low).  
In case of non-axisymmetric  boundary conditions at the top of the fluid core,
the azimuthal heat flux becomes important.
The azimuthal heat flux is always negative and increases towards
the equator. 
This leads to a latitude dependent westward transport of
heat which should interact with the constraints given by the inhomogeneous
thermal conditions at the core mantle boundary.

Most considerations within the presented work rely on the assumption of a
dominant toroidal field within the fluid core. 
This does not have to be necessarily the case.
Due to the insulating mantle the toroidal component of the Earth's magnetic field is hidden from any observations. 
Indeed, most of the authors favor a geodynamo-model of $\alpha\Om$-type (Fearn 1998)
which is characterized by a dominant azimuthal field.
However, estimations of the toroidal field strength from the extrapolation
of the dc electric potential near the top of the Earth's mantle to the core
mantle boundary result in comparable strength of the toroidal and the poloidal
field components (Levy \& Pearce, 1991).
Furthermore, recent simulations of Christensen \& Aubert (2006) that cover a
very large parameter space, yield solutions without significant differential
rotation where poloidal and
toroidal field strength are comparable ($\alpha^2$-Dynamo). 

Independent from the orientation of the dominant field component, the
direction of the horizontal components of the heat flux (towards
the poles and westwards, i.e. against the direction of the rotation) is a
consequence of the rotation and not caused by the magnetic field which only
enhances or reduces the magnitude of the heat flux.
Therefore the horizontal components of the heat flux should show a similar sign even in
case of comparable  toroidal and poloidal field strength.  

The circumstances within a rotating spherical shell are more
complicated as they appear in the presented Cartesian model where relevant
characteristics have not been concerned for simplicity.
In particular, the  combined effects of all three components of the turbulent
heat flux cannot be inferred from a simple local model so that the consequences of the common behavior could only be examined by a global
simulation.
However, the results should at least qualitatively resemble the combined effects of the small scale
fluctuations on large scale fields and comparable effects are expected to be present in
spherical geometries. To what extend the reported results for
the heat flux will endure in simulations in spherical geometry, where also
possible larger scale flows will occur, again can only be clarified in highly
resolved global simulations.

\bibliographystyle{gji}

%
\end{document}